\documentclass[journal,twocolumn]{IEEEtran}
\usepackage{amsmath,amsfonts}
\usepackage{algorithmic}
\usepackage{algorithm}
\usepackage{array}
\usepackage{textcomp}
\usepackage{stfloats}
\usepackage{subfig}
\usepackage{url}
\usepackage{verbatim}
\usepackage{graphicx}
\usepackage{cite}
\usepackage{multirow}
\usepackage{color}
\usepackage{graphicx}

\begin{document}
	
	\title{A Hypernetwork Based Framework for Non-Stationary Channel Prediction} 
	
	\author{Guanzhang Liu, Zhengyang Hu, Lei Wang, Hongying Zhang, Jiang Xue,~\IEEEmembership{Senior Member,~IEEE}, and Michail Matthaiou,~\IEEEmembership{Fellow,~IEEE}
		
		\thanks{Copyright (c) 2015 IEEE. Personal use of this material is permitted. However, permission to use this material for any other purposes must be obtained from the IEEE by sending a request to pubs-permissions@ieee.org.}
		
		\thanks{Guanzhang Liu, Zhengyang Hu, Lei Wang and Hongying Zhang are with the School of Mathematics and Statistics, Xi'an Jiaotong University, Xi'an 710049, China (e-mail: lgzh97@stu.xjtu.edu.cn, hzyxjtu@stu.xjtu.edu.cn, wl\_simple@stu.xjtu.edu.cn, zhyemily@mail.xjtu.edu.cn).}% <-this % stops a space
		
		\thanks{Jiang Xue is with the School of Mathematics and Statistics, Xi'an Jiaotong University, Xi'an 710049, China, and also with the Peng Cheng Laboratory, Shenzhen, Guangdong 518055, China, and also with the Pazhou Laboratory (Huangpu), Guangzhou, Guangdong 510555, China (e-mail: x.jiang@xjtu.edu.cn).}%
		
		\thanks{Michail Matthaiou is with the Centre for Wireless Innovation (CWI), Queen’s University Belfast, BT3 9DT Belfast, U.K. (e-mail: m.matthaiou@qub.ac.uk).}% <-this % stops a space
		
		\thanks{The corresponding author is Jiang Xue. The work of Jiang Xue and Guanzhang Liu was supported in part by the National Key R$\&$D Program of China under Grant 2020YFA0713900, in part by the major key project of Peng Cheng Laboratory under grant PCL2023AS1-2, and in part by the Joint Key Project of Universities Shaanxi (S2023-YF-GXZD-0022). The work of Hongying Zhang was supported in part by the National Natural Science Foundation of China (Nos. 12171386). The work of Michail Matthaiou was supported in part by the European Research Council (ERC) under the European Union’s Horizon 2020 research and innovation programme (grant agreement No. 101001331).}}
	
	\maketitle
	
	\begin{abstract}
		In order to break through the development bottleneck of modern wireless communication networks, a critical issue is the out-of-date channel state information (CSI) in high mobility scenarios. In general, non-stationary CSI has statistical properties which vary with time, implying that the data distribution changes continuously over time. This temporal distribution shift behavior undermines the accurate channel prediction and it is still an open problem in the related literature. In this paper, a hypernetwork based framework is proposed for non-stationary channel prediction. The framework aims to dynamically update the neural network (NN) parameters as the wireless channel changes to automatically adapt to various input CSI distributions. Based on this framework, we focus on low-complexity hypernetwork design and present a deep learning (DL) based channel prediction method, termed as LPCNet, which improves the CSI prediction accuracy with acceptable complexity. Moreover, to maximize the achievable downlink  spectral efficiency (SE), a joint channel prediction and beamforming (BF) method is developed, termed as JLPCNet, which seeks to predict the BF vector. Our numerical results showcase the effectiveness and flexibility of the proposed framework, and demonstrate the superior performance of LPCNet and JLPCNet in various scenarios for fixed and varying user speeds.
	\end{abstract}
	
	\section{Introduction}
	
	As 5G networks are being deployed worldwide, researchers from academia and industry have started looking into the development of beyond 5G and 6G networks \cite{Matthaiou:COMMag:2021}. In the evolution of wireless communication networks, the growing number of antennas per $\textrm{km}^2$ is a prominent hallmark, thereby greatly improving energy and spectral efficiency \cite{zhang2020prospective,mMIMO,EE}. To fully avail of the advantages of the large number of antennas, it is vital to acquire timely and accurate channel state information (CSI) at the base station (BS). However, either the feedback delay in the frequency division duplex (FDD) mode or the processing delay in the time division duplex (TDD) mode can lead to outdated CSI, also known as the channel aging phenomenon \cite{Aging}, which negatively impacts the system performance of wireless communication networks. More importantly, the mobility of users (UEs) will further compromise the system performance, especially in high mobility scenarios \cite{Hmobility,TimeCorr,TimeCorr2}.
	
	Fortunately, during the propagation of electromagnetic (EM) waves, the scattering environment is the same in a fixed region, resulting in correlations across CSI in the temporal domain \cite{TimeCorr1}. Therefore, to overcome the channel aging problem, a natural way is to predict CSI in the future, a technique known as channel prediction. In the literature, numerous channel prediction approaches have been proposed, e.g., the parametric radio channel (PRC) model based predictor, the autoregressive (AR) model based predictor, and the recently emerging deep learning (DL) model based predictor.
	
	\subsection{Prior State-of-the-Art}
	
	In general, the PRC model based predictor estimates the propagation parameters (e.g., Doppler and amplitude) in the channel model and predicts CSI via the estimated parameters \cite{PRC}. This kind of approach is commonly based on the quasi-static assumption, which means that the propagation parameters should be time-invariant. Depending on the parameter estimation, the PRC model based predictor includes the maximum likelihood based method \cite{ML}, the subspace based method \cite{subsp1}, the compressed sensing based method \cite{CS}, etc. The authors of \cite{HF} proposed a more realistic approach by considering the time-variant path delays and proposed a super-resolution channel prediction method. Nevertheless, in high mobility scenarios, to achieve accurate channel prediction, the BS needs to perform the parameter estimation frequently, which inherently entails high computational complexity.
	
	Without extraction of the propagation parameters, the AR model based predictors directly predict the entire CSI, and have a lower computational complexity than the PRC model based predictors. Specifically, the AR model maps the historical CSI to the future linearly, which requires the wide-sense stationary (WSS) assumption to hold \cite{AR}. In \cite{CorrAR}, the authors exploited the spatial correlation of CSI and derived two AR based predictors with different data selection strategies. In addition, there have also been some works that focus on utilizing the characterization of CSI in the angle-delay domain, such as the Prony-based angular-delay domain (PAD) predictor \cite{PAD} and the spatio-temporal autoregressive (ST-AR) predictor \cite{STAR}. However, the inherent linear and WSS constraints limit their modeling capabilities in realistic high mobility scenarios, in which wireless channels may exhibit nonlinear and non-stationary dynamics. Motivated by the nonlinear approximation power of higher-order differential equation (HODE), the authors of \cite{HODE} incorporated genetic programming (GP) and HODE, and presented a novel nonlinear channel prediction scheme.
	
	Recently, deep learning (DL) has attracted unprecedented research attention in the fields of computer vision and natural language processing, demonstrating strong nonlinear learning capabilities. In wireless communication networks, DL is also driving the development of 5G and beyond \cite{DL5G}. Interestingly, most of the algorithms in the physical (PHY) layer have been redesigned by DL models, such as channel estimation \cite{CE}, signal detection \cite{SD}, precoding \cite{precoding} and channel feedback \cite{feedback,lora}. Not surprisingly, there are also some attempts to apply DL models to the channel prediction problem. 
	
	In \cite{MLP}, the authors proposed a multi-layer perception (MLP) based predictor and a vector Kalman filter (VKF) based predictor. Their results showed that the MLP based predictor is comparable with the VKF based predictor in terms of the CSI prediction accuracy, and has a much lower computational complexity once trained. Later, in \cite{bayes}, the MLP network was combined with Bayesian learning for hyperparameter optimization in channel prediction. Motivated by the great potential of the recurrent neural networks (RNNs) and their variants in the area of series modeling, a RNN based predictor \cite{rnn} and a long short-term memory (LSTM) based predictor \cite{lstm} were respectively proposed. To describe the continuous physics process of wireless channels, based on the RNNs and ordinary differential equation (ODEs), the authors of \cite{ode} developed a novel predictor with higher interpretability. Considering the problem of insufficient datasets, the authors of \cite{gan} designed a generative adversarial network (GAN) and LSTM based channel prediction framework. Moreover, by visualizing the CSI matrix as a two-dimensional image, in \cite{cnnlstm}, the authors employed convolutional neural network (CNN) and RNN in cascade to extract spatial and temporal features, respectively. In \cite{clstm}, convolutional LSTM (ConvLSTM) was used to jointly exploit the spatio-temporal correlations of CSI for the high-speed train (HST) channel prediction. Furthermore, the work in \cite{drnn} demonstrated that a better performance can be achieved by predicting the difference between CSI. Instead of discriminating CSI directly, the hidden states of ConvLSTM cells were processed by the difference operator, whilst the attention mechanism was used to implement the feature refinement for multi-step channel prediction \cite{stnnn}. Since the recurrent structure may lead to the accumulation of the CSI prediction error, a transformer based scheme was proposed to predict CSI sequence in parallel \cite{transformer}. With the aid of additional position information of the UE, a Cosine-Gaussian Radial Basis Function network (C-GRBFnet) was introduced to improve the accuracy of channel prediction in \cite{GRBF}.
	
	Although the existing DL based predictors yield significant performance gains, the non-stationary nature of CSI in modern wireless communication networks should be carefully investigated. The non-stationarity of the wireless channel causes its statistical properties, such as mean and variance, to change temporally, which implies that the data distribution shifts continuously over time. Not only the distribution of the test data is different from the training data, but also the input sequences to the DL models may follow various underlying distributions. This temporal distribution shift problem is ignored by the aforementioned DL based methods and compromises the system performance \cite{revin}. In addition, the purpose of channel prediction is to acquire the specific values of CSI by reducing the mean square error (MSE), whilst it is further utilized for the beamforming (BF) design. From a divide-and-conquer perspective \cite{funcom}, channel prediction and BF are typically treated as two independent blocks, and the most significant performance metrics, i.e., achievable downlink spectral efficiency (SE), are affected by the BF vector. However, a lower MSE of channel prediction does not necessarily lead to a higher achievable downlink SE.
	
	\subsection{Main Contributions}
	
	In this paper, we propose a hypernetwork based framework for non-stationary channel prediction to tackle the temporal distribution shift problem brought by the CSI non-stationarity. In the proposed framework, the parameters of the prediction neural network (NN) branch are adjusted by the hypernetwork branch for each input CSI sequence, thereby realizing instance-wise adaptation. Under the proposed framework, we focus on low-complexity hypernetwork design and develop a DL based channel prediction method, dubbed LPCNet. By taking the non-stationarity of wireless channels into account, LPCNet greatly improves the CSI prediction accuracy with a little complexity increase. Furthermore, inspired by the end-to-end design of \cite{e2e,Guo,LiuAn}, we present a DL based joint channel prediction and BF method, dubbed JLPCNet, which aims to maximize the BF performance rather than the CSI prediction accuracy. 
	
	The major contributions of this paper are summarized as follows:
	\begin{itemize}
		\item[$\bullet$]After analyzing the effect of the CSI non-stationarity on the channel prediction, we propose a novel hypernetwork based channel prediction framework for non-stationary wireless channels. Considering the flexibility and complexity of the framework, the output of the adjustor is only utilized to adjust the parameters of the predictor. Once the offline training is completed, the parameters in the prediction NN branch can be updated over time for each input CSI sequence without retraining.
		\item[$\bullet$]We design a new DL based channel prediction method, namely, LPCNet, in which a dynamic linear layer is employed to adapt to different input CSI distributions. The parameters of the dynamic linear layer are optimized by two parameter sharing light MLP networks. To the best of our knowledge, this work is the first to address the temporal distribution shift problem in channel prediction. 
		\item[$\bullet$]For the purpose of improving the BF performance rather than the CSI prediction accuracy, we focus on predicting the BF vector with specific direction. Without acquiring the future CSI, JLPCNet directly predicts the future BF vector by inputting the past CSI sequence. The cosine similarity based loss function is built to optimize the parameters of JLPCNet, which concentrates on the directional information of CSI.
		\item[$\bullet$]We evaluate the proposed methods on the QuaDRiGa channel generator \cite{Quadriga} through extensive comparisons. Numerical results demonstrate the effectiveness and flexibility of the proposed framework, and that LPCNet has an excellent CSI prediction accuracy in different scenarios for fixed and varying UE speeds. Moreover, we show that JLPCNet achieves a superior BF performance than its comparative benchmarks. 
	\end{itemize}
	
	\subsection{Organization and Notations}
	
	The rest of this paper is organized as follows: In Section \uppercase\expandafter{\romannumeral2}, we introduce the channel model and the signal transmission model. In Section \uppercase\expandafter{\romannumeral3}, we present the motivation, the problem formulation and the hypernetwork based framework for non-stationary channel prediction. Section \uppercase\expandafter{\romannumeral4} describes the details of LPCNet and JLPCNet. The numerical settings and results are shown in Section \uppercase\expandafter{\romannumeral5}. Finally, Section \uppercase\expandafter{\romannumeral6} concludes this paper.
	
	\textbf{Notations}: Throughout the paper, we use bold uppercase letters to denote matrices, bold lowercase letters to denote vectors, and non-bold letters to denote scalars; $|\cdot|$ stands for the element-wise absolute value; $\Vert \cdot \Vert_2$ is the Euclidean norm; $(\cdot)^T$ and $(\cdot)^H$ represent the transpose and conjugate transpose. The complex number field is represented by $\mathbb{C}$ and the real number field is represented by $\mathbb{R}$. The symbol $\circ$ represents the Hadamard product; $\mathbb{E}\{\cdot\}$ represents the expectation operation. Finally, $\mathcal{C}\mathcal{N}(\mu,\sigma^{2})$ denotes the complex Gaussian distribution with mean $\mu$ and variance $\sigma^{2}$.
	
	\section{System Model}
	
	In this section, we present the time-varying channel model and the signal transmission model in detail. 
	
	\subsection{Time-Varying Channel Model}
	
	We consider a single-cell downlink communication system, where a fixed BS serves a mobile UE at frequency $f$. The corresponding wavelength is denoted as $\lambda$. The BS is equipped with a dual polarized, uniform planar array (UPA), comprising $N_l$ antennas in a column and $N_r$ antennas in a row. The distance between two adjacent antennas in the horizontal or vertical direction is half of the wavelength, i.e., $d=\lambda/2$. The UE is equipped with an omni-directional antenna. Note that $N_b$ denotes the number of BS antennas. It is clear that $N_b=2N_lN_r$. 
	
	We adopt a 3-D time-varying multipath channel model from the 3GPP technical report \cite{CM}. We denote the channel frequency response between the UE and the $b$-th transmit antenna element at time $t$ as $h_b(t)$, which is modeled as 
	\begin{equation} \label{channelmodel}
	h_{b}(t) = \sum_{l=1}^{L}\alpha_le^{j2\pi\frac{\hat{\mathbf r}_{\textrm{tx},l}^{T} \overline{\mathbf d}_{\textrm{tx},b}}{\lambda}}e^{j2\pi\frac{\hat{\mathbf r}_{\textrm{rx},l}^{T} \overline{\mathbf v}} {\lambda}t}e^{-j2\pi f \tau_l},
	\end{equation}
	where $\overline{\mathbf v}$ represents the UE velocity vector with speed $v$, travel azimuth angle $\phi_v$ and travel elevation angle $\theta_v$, which is expressed as 
	\begin{equation}\label{v}
	\overline{\mathbf v} = v        
	\begin{bmatrix}
	\sin\theta_v\cos\phi_v,\sin\theta_v\sin\phi_v,\cos\theta_v
	\end{bmatrix}^T.
	\end{equation}
	Additionally, $\hat{\mathbf r}_{\textrm{rx},l}$ and $\hat{\mathbf r}_{\textrm{tx},l}$ represent the spherical unit vectors of the UE and the BS corresponding to the $l$-th path, which are described as 
	\begin{equation}
	\hat{\mathbf r}_{\textrm{rx},l} = 
	\begin{bmatrix}
	\sin\theta_{l,\textrm{EOA}}\cos\phi_{l,\textrm{AOA}}\\
	\sin\theta_{l,\textrm{EOA}}\sin\phi_{l,\textrm{AOA}}\\
	\cos\theta_{l,\textrm{EOA}}
	\end{bmatrix},
	\end{equation}
	and 
	\begin{equation}
	\hat{\mathbf r}_{\textrm{tx},l} = 
	\begin{bmatrix}
	\sin\theta_{l,\textrm{EOD}}\cos\phi_{l,\textrm{AOD}}\\
	\sin\theta_{l,\textrm{EOD}}\sin\phi_{l,\textrm{AOD}}\\
	\cos\theta_{l,\textrm{EOD}}
	\end{bmatrix}.
	\end{equation}
	For ease of reading, the definitions of other parameters in the channel model are listed in Table \ref{channelparams}. 
	
	\begin{table}[t]
		\centering
		\caption{The definition of parameters in the channel model.}\label{channelparams}
		\renewcommand\arraystretch{1.3}
		\begin{tabular}{c|c}
			\hline
			Parameter & Definition \\
			\hline\hline
			$L$ &  number of paths \\
			\hline
			$\alpha_l$ &  complex attenuation coefficient of the $l$-th path \\
			\hline
			$\phi_{l,\textrm{AOA}}$ &  azimuth angle of arrival (AOA) of the $l$-th path \\
			\hline
			$\theta_{l,\textrm{EOA}}$ &  elevation angle of arrival (EOA) of the $l$-th path \\
			\hline
			$\phi_{l,\textrm{AOD}}$ &  azimuth angle of departure (AOD) of the $l$-th path \\
			\hline
			$\theta_{l,\textrm{EOD}}$ &  elevation angle of departure (EOD) of the $l$-th path \\
			\hline
			$\tau_l$ &  propagation delay of the $l$-th path \\
			\hline
			$\overline{\mathbf d}_{\textrm{tx},b}$ & location vector of the  $b$-th transmit antenna element \\
			\hline
		\end{tabular}
	\end{table}
	
	\subsection{Signal Transmission Model}
	
	Let $s(t)\sim \mathcal{C}\mathcal{N}(0,1)$ denote the data symbol from the BS to be sent to the UE at time $t$, and let $\mathbf{w}(t)\in\mathbb{C}^{N_b}$ denote the linear BF vector, which satisfies the BS power constraint, i.e., $|\mathbf{w}(t)^H\mathbf{w}(t)| \leq P$. The corresponding received signal in the data transmission phase can be written as 
	\begin{equation}\label{receivedsignal}
	y(t) = \mathbf{h}(t)^T\mathbf{w}(t)s(t) + n(t),
	\end{equation}
	where $\mathbf{h}(t)\in\mathbb{C}^{N_b}$ is the downlink channel vector between the BS and the UE at time $t$ and $n(t) \sim \mathcal{C}\mathcal{N}(0,\sigma^{2}_n)$ is the independent additive white Gaussian noise (AWGN). Given the received signal model in (\ref{receivedsignal}), the achievable downlink SE is expressed as follows 
	\begin{equation}\label{rate}
	R=\textrm{log}_2\left(1+\frac{|\mathbf{h}(t)^T\mathbf{w}(t)|^2}{\sigma_n^2}\right).
	\end{equation}
	
	In order to design the optimal BF vector, it is a common assumption to have access to perfect instantaneous CSI at the BS. However, due to variations of the wireless channel caused by the movement of the UE and the processing delay, the acquisition of $\mathbf{w}(t)$ may be based on the outdated CSI, such as $\mathbf{h}(t-1)$. In other words, during the actual downlink data transmission phase, the wireless channel has already changed. The outdated CSI results unavoidably render the BF strategy outdated, which, in turn, cannot ensure a reliable data transmission. For example, the authors of \cite{PAD} reported that the performance drops up to 50$\%$, when the UE speed is increased from 3 km/h to 30 km/h. By exploiting the inherent temporal correlation of wireless channels, predicting CSI in the future is utilized to address this channel aging problem.
	
	\section{Non-Stationary Channel Prediction Framework}
	
	In this section, we analyze the CSI non-stationarity using time series analysis tools, reformulate the channel prediction problem on the basis of the non-stationary property, and introduce a novel framework based on the hypernetwork to deal with the non-stationary channel prediction problem. 
	
	\subsection{CSI Non-Stationarity}
	
	\begin{figure}[t]
		\centerline{\includegraphics[width=3.5in]{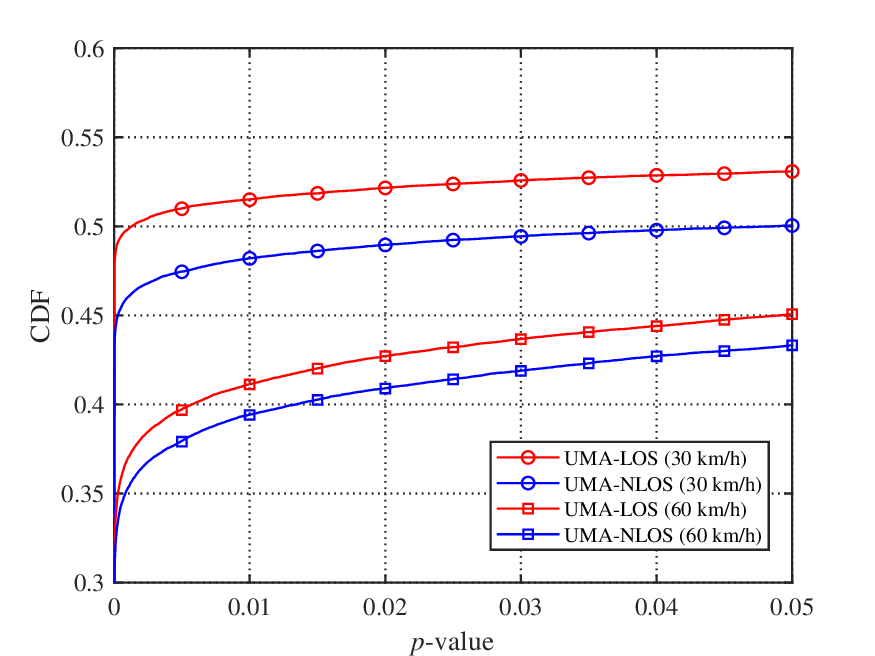}}
		\caption{The CDF curves of the $p$-value under the urban macro-line-of-sight (UMA-LOS) and UMA-non-line-of-sight (UMA-NLOS) scenarios for different UE speeds. The time-varying channels are split into several segments based on a sliding window, and the ADF test is applied to the CSI sequence in each segment.}
		\label{Fig1}
	\end{figure}
	
	The WSS assumption of wireless channels is widely used for the analysis of CSI statistics and the design of algorithms in the PHY layer. However, some recent measurement results \cite{NS,NSM,NSM1,NCM} articulated that the WSS assumption is not always satisfied in practice, especially in HST and vehicle-to-everything (V2X) channels. The high mobility and plethora of multi-path components in modern wireless communication networks are all contributing factors to the non-stationary behavior of wireless channels. From a time series analysis perspective, we employ the augmented Dickey-Fuller (ADF) test\footnote{The null hypothesis of the ADF test is that there is a unit root (i.e., non-stationarity), with the alternative that there is no unit root (i.e., stationarity).} \cite{adf} to validate the CSI non-stationarity. Figure \ref{Fig1} presents the cumulative distribution function (CDF) curves of the $p$-value.\footnote{The smaller the $p$-value, the greater the statistical incompatibility of the data with the null hypothesis. The $p$-value is used for comparison with a significance level. If the $p$-value is above a significance level, then we cannot reject that there is a unit root. The typical value of the significance level is 0.01 or 0.05.} A detailed description of the propagation environment is given in Section \uppercase\expandafter{\romannumeral5}-A. From the figure, it is seen that an increase in the movement speed significantly aggravates the CSI non-stationarity in both the LOS and NLOS environments. Additionally, we can observe that the NLOS environment with far more scatterers clearly exhibits stronger non-stationarity than the LOS environment for a certain UE speed. Depending on the existence of the non-stationarity in modern wireless channels, there are a few studies that have started to concentrate on the design of non-stationary property based algorithms \cite{NAD}, motivating us to implement non-stationary channel prediction.
	
	\subsection{Problem Formulation}
	
	Generally, channel prediction is carried out using a sliding window pattern. By constructing a reasonable function map $F$ to model the relationship between the past CSI sequence and the future CSI, we implement the channel prediction within the window. Once the architecture $\mathcal{F}$ and the parameter set $\Theta$ of the function map $F$ are determined, they stay fixed during the online interference/prediction period. Nevertheless, the non-stationary property of wireless channels creates the discrepancies between the distributions of the CSI sequences in different windows, which poses a challenge to the adaptability of the prediction model and is the most critical issue in this paper. To overcome the substantial distribution discrepancies between the CSI sequences in different windows, the AR model commonly requires to update their parameters over time. The result in \cite{STAR} shows that the CSI prediction error of the AR model gradually increases with the parameter update period. Inspired by this, we adapt the NN parameters to different input CSI sequences to solve the temporal distribution shift problem in non-stationary channel prediction. In contrast to the existing DL based prediction models with a fixed architecture $\mathcal{F}$ and the static parameter set $\Theta$, DL based prediction models with a dynamic parameter set, can be expressed as
	\begin{equation}\label{dynamicnn}
	\hat{\mathbf{h}}_j(t) = 
	\mathcal{F}\left(\mathbf{X}_j(t);\Theta_j\right),
	\end{equation}
	where $j$ denotes the index of the $j$-th sliding window, corresponding to a specific input CSI sequence\footnote{In this paper, we assume that the past CSI sequence has been perfectly obtained at the BS side.} $\mathbf{X}_j(t)=[\mathbf{h}_j(t-1),\mathbf{h}_j(t-2),\ldots,\mathbf{h}_j(t-K)]$ and a predicted CSI vector $\hat{\mathbf{h}}_j(t)$, $K$ is the length of the past CSI sequence, while $\Theta_j$ denotes the instance-dependent NN parameter set. The function map described in (\ref{dynamicnn}) no longer treats different inputs indiscriminately as before, but adapts to different inputs dynamically. That is to say, we leverage the instance-dependent prediction function to take the instance-wise distribution discrepancy into account. In addition, the input CSI sequence $\mathbf{X}_j(t)$ within the window can be non-stationary. Although DL models make no explicit assumptions about the temporal dynamics of its input, we intend to mitigate the non-stationarity of the input CSI sequence to DL models.\footnote{Note that although the non-stationarity of the CSI sequence within the window is mitigated, the distribution discrepancies between the CSI sequences in different windows cannot be overlooked in high mobility scenarios.} Our approach is consistent with some prior works (e.g., \cite{drnn,stnnn}), which have demonstrated that processing the non-stationarity of the input CSI sequence is beneficial for enhancing the CSI prediction accuracy in practice. 
	
	Motivated by the above mentioned discussion on the effects of the CSI non-stationarity on channel prediction, a novel channel prediction framework is developed in Section \uppercase\expandafter{\romannumeral3}-C, which is suitable for non-stationary wireless channels.
	
	\subsection{Hypernetwork Based Framework}
	
	For the purpose of realizing the NN parameters dependent on the specific input, we leverage the idea of hypernetwork, which was firstly introduced in \cite{hypernet} in the field  of image recognition and neural machine translation. In the space of wireless communications, hypernetworks have been applied to channel feedback \cite{hyperfeedback} and signal detection \cite{hypermm}, bringing positive performance gains. Given a new input instance, the goal of a hypernetwork is to use an extra NN to predict the parameters of the task NN. As a part of the entire NN architecture, this extra NN joins the end-to-end training by back-propagation, and thus completes the fast adaptation without retraining. Motivated by this, we design a novel framework for non-stationary channel prediction, consisting of a prediction NN branch and a hypernetwork branch, as shown in Fig. \ref{Fig2}. The prediction NN branch is used to acquire the future CSI by inputting the past CSI sequence within the window, while the hypernetwork branch produces a specific parameter adjustment for the prediction NN branch when a new input CSI sequence comes from the sliding window. More specifically, the proposed framework consists of four modules with the respective roles, described as follows:
	
	\begin{figure*}[t]
		\centerline{\includegraphics[height=2.2in]{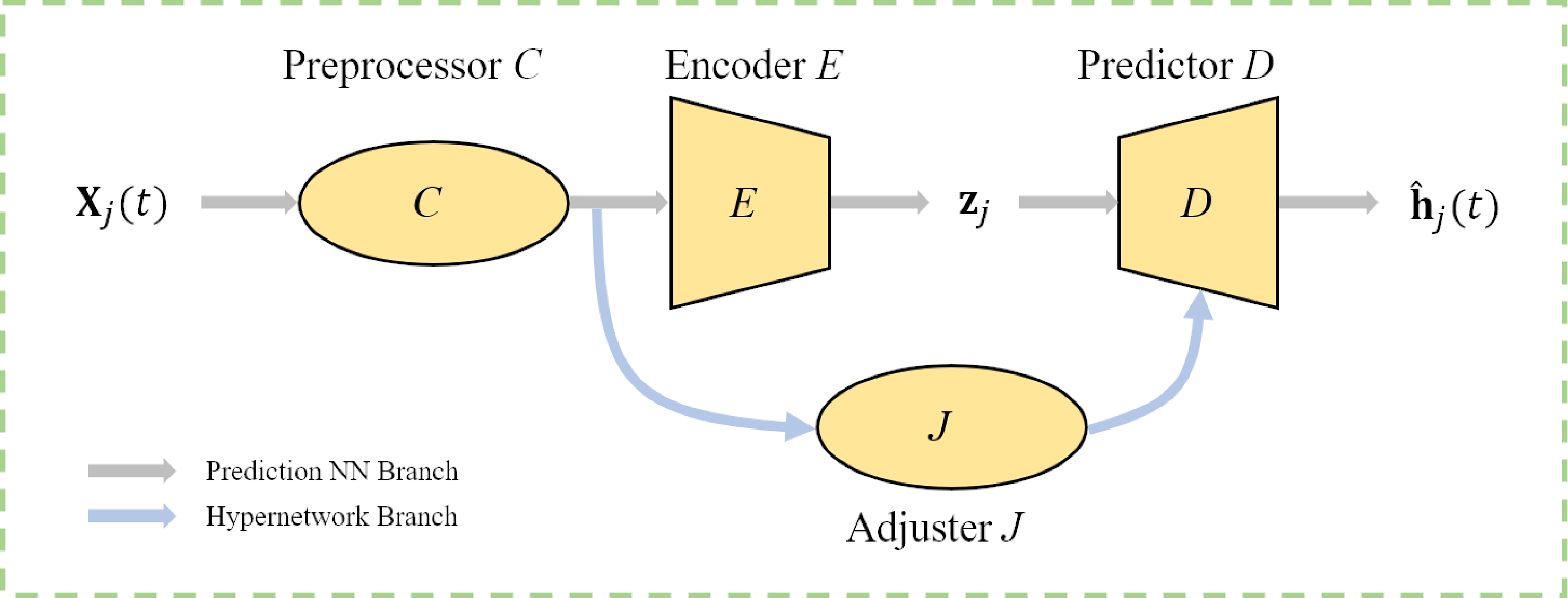}}
		\caption{The proposed hypernetwork based framework for non-stationary channel prediction.}
		\label{Fig2}
	\end{figure*}
	
	\subsubsection{Preprocessor $C$}
	The goal of the preprocessor $C$ is to alleviate the non-stationary dynamics of the input CSI sequence to the prediction model, and provides an approximately stationary sequence. The preprocessor $C$ is helpful for modeling the temporal dependencies and achieving the effective feature extraction for the subsequent modules. 
	\subsubsection{Encoder $E$}
	The encoder $E$ is essential for DL based prediction models, and its design is a main focus in the existing literature (e.g., \cite{rnn,lstm,ode,cnnlstm,clstm,stnnn}). By extracting the useful features from the input, the encoder $E$ generates a feature representation vector $\mathbf{z}_j$. The powerful deep time series models, such as RNN and LSTM, are mainly considered in the design of the encoder $E$.
	\subsubsection{Predictor $D$}
	The predictor $D$ aims to learn a mapping from the latent space to the label space. Concretely, the feature representation vector $\mathbf{z}_j$ is mapped to the future CSI vector $\hat{\mathbf{h}}_j(t)$ based on the predictor $D$, which is commonly composed of a simple linear layer or convolution layer.
	\subsubsection{Adjuster $J$}
	To tackle the distribution discrepancies between the input sequences caused by the CSI non-stationarity, the adjuster $J$ is the core module in the proposed framework to realize the purpose of the dynamic parameter set. For a given CSI sequence, denoted by $\mathbf{X}_j(t)$, the parameters of the predictor $D$ are instance-dependent, which are adjusted by the output of the adjuster $J$. Once the end-to-end training is completed, the parameters of the adjuster $J$ are fixed, while the parameters in the predictor $D$ update along with the input CSI sequence temporally.
	
	In the proposed framework, only the parameters of the predictor $D$ are dynamically adjusted, while the encoder $E$ of the prediction NN branch is not optimized by the adjuster $J$. There are two reasons behind this approach. On one hand, employing the hypernetwork branch to adjust the NN parameters will inevitably bring additional computation. The NN architecture of the encoder is usually complex and contains numerous parameters. To update all parameters in the encoder with the respective adjustments, the computational complexity may be heavy. In contrast, the predictor is relatively simple and can improve the performance whenever optimized by the adjuster. On the other hand, such a design also allows the proposed framework to be flexible, which means that it is model-agnostic and can be applied to various channel prediction scenarios. In Section \uppercase\expandafter{\romannumeral5}-B, we will show that the CSI prediction accuracy of both recursive and non-recursive models can be boosted by incorporating them into our framework. Although a single-antenna UE is considered in the system model, it is straightforward to apply the proposed framework to UEs with multiple antennas. For example, the vectorization form of the CSI matrix or popular spatio-temporal models, such as ConvLSTM \cite{stnnn}, can be integrated into the proposed framework.
	
	\section{Design of LPCNet and JLPCNet}
	
	In this section, we first present a new DL based prediction method with dynamic parameters, namely, LPCNet, to implement the non-stationary channel prediction in (\ref{dynamicnn}). Then, in order to maximize the achievable downlink SE, we further provide a DL based joint channel prediction and BF design, namely, JLPCNet.
	
	\subsection{LPCNet}
	
	In this subsection, the details of LPCNet are described, including the NN architecture, training procedure and complexity analysis. 
	
	Denote $\{\mathbf{h}_{j,t}\}_{t=1}^{K}$ as the input CSI sequence through the $j$-th sliding window. Due to the requirement of the real-valued input, each CSI vector $\mathbf{h}_{j,t}$ is transformed to its real-valued form with size $2N_b$ after splitting the real and imaginary parts. Hereafter, we consider $\{\bar{\mathbf{h}}_{j,t}\}_{t=1}^{K}$ as the real-valued input CSI sequence to LPCNet.
	
	\begin{figure}[t]
		\centerline{\includegraphics[width=3.0in]{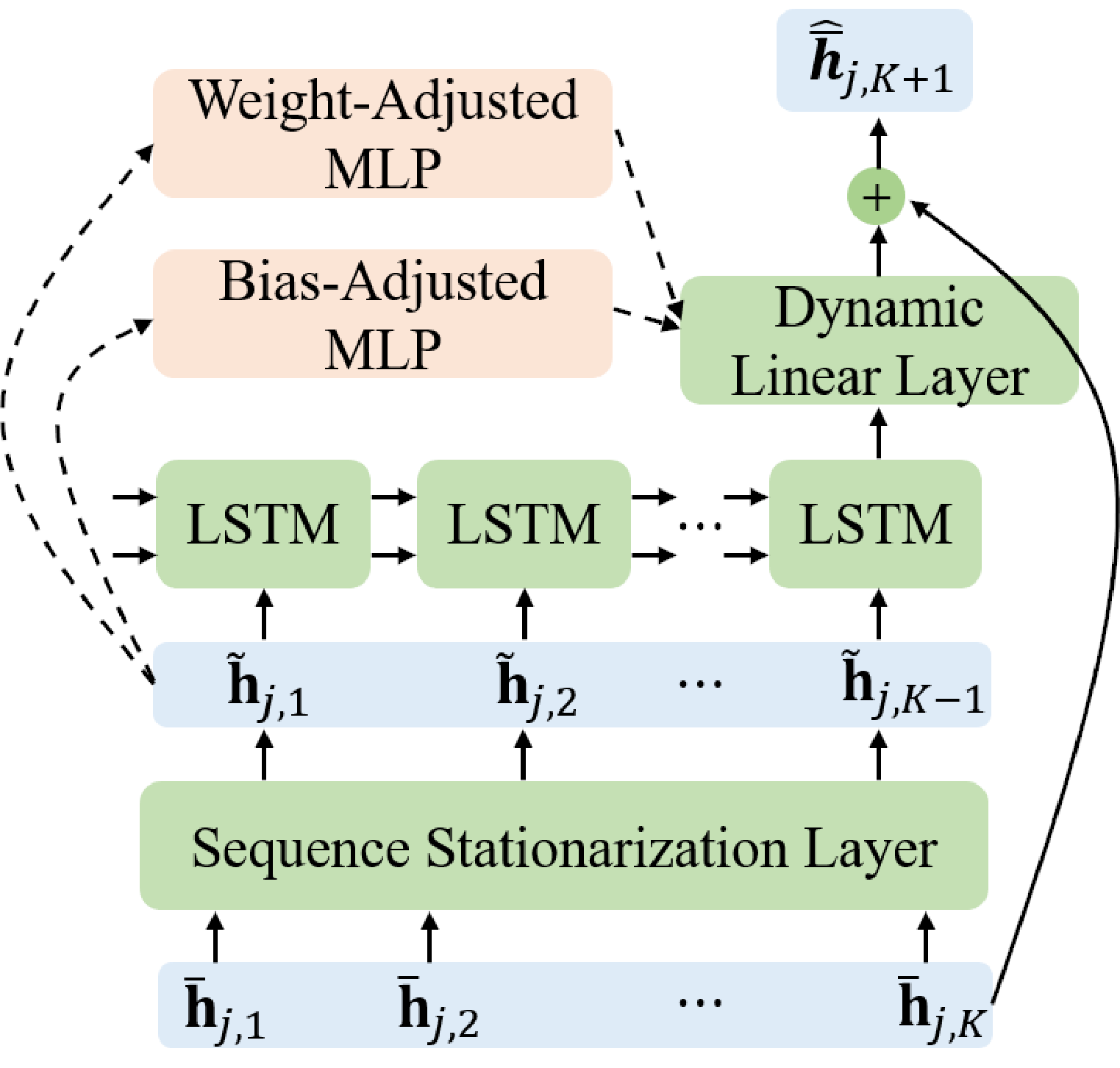}}
		\caption{Illustration of the NN architecture and forward procedure of LPCNet.}
		\label{Fig3}
	\end{figure}
	
	\subsubsection{Architecture of LPCNet}
	
	Based on the proposed framework for solving the non-stationary channel prediction problem, we design a DL based method for learning to predict CSI, called LPCNet, which implements the instance-dependent channel prediction. Figure \ref{Fig3} shows the overall architecture of LPCNet, in which the solid lines indicate the forward procedure of the prediction branch to output the future CSI vector while the dashed lines indicate the forward procedure of the hypernetwork branch for the purpose of dynamic parameters. First of all, a sequence stationarization layer is designed as the preprocessor $C$ to mitigate the non-stationarity of the input CSI sequence $\{\bar{\mathbf{h}}_{j,t}\}_{t=1}^{K}$. As a simple and effective approach, the first-order difference operator \cite{ARIMA} is adopted and a new sequence $\{\tilde{\mathbf{h}}_{j,t}\}_{t=1}^{K-1}$ is derived by
	\begin{equation}\label{diff}
	\tilde{\mathbf{h}}_{j,t}=\bar{\mathbf{h}}_{j,t+1}-\bar{\mathbf{h}}_{j,t}.
	\end{equation}
	We now note that a LSTM circumvents the long-term dependence problem in general RNNs very well, and has shown an excellent CSI prediction accuracy \cite{lstm,gan}. Since our objective is to model the distribution discrepancies between the input CSI sequences through the design of the hypetnetwork branch, the encoder $E$ utilizes the LSTM cell to extract features from the sequence $\{\tilde{\mathbf{h}}_{j,t}\}_{t=1}^{K-1}$. Its recursive process can be summarized as 
	\begin{equation}\label{lstm}
	\{\mathbf c_{j,t},\mathbf z_{j,t}\}=\text{LSTM}(\mathbf c_{j,t-1},\mathbf z_{j,t-1},\tilde{\mathbf{h}}_{j,t};\Omega_t),
	\end{equation}
	where $\mathbf c_{j,t}$($\mathbf c_{j,t-1}$) is called the cell state and $\mathbf z_{j,t}$($\mathbf z_{j,t-1}$) is called the hidden state, $\mathbf c_{j,t}$ and $\mathbf z_{j,t}$ are the output at time $t$, $\mathbf c_{j,t-1}$, $\mathbf z_{j,t-1}$ and $\tilde{\mathbf{h}}_{j,t}$ are the input at time $t$. Also, $\Omega_t$ is the parameter set of LSTM, which is shared along time, i.e., $\Omega_t=\Omega$. Denote the number of neurons of LSTM as $N_z$. For brevity, we no longer elaborate the principle of LSTM, and more details about LSTM can be found in \cite{olstm}. At last, a dynamic linear layer is applied on the LSTM output $\mathbf z_{j,K-1}$, and the original information $\bar{\mathbf{h}}_{j,K}$ is added to predict the specific values of the future CSI vector $\bar{\mathbf{h}}_{j,K+1}$. The predicted CSI in the label space can be obtained by 
	\begin{equation}\label{dlinear}
	\hat{\bar{\mathbf{h}}}_{j,K+1}= (\mathbf W^\text{a}_j\circ \mathbf W)\mathbf z_{j,K-1}+\mathbf b^\text{a}_j \circ \mathbf b + \bar{\mathbf{h}}_{j,K},
	\end{equation}
	where $\mathbf W\in\mathbb{R}^{2N_b\times N_z}$ and $\mathbf b\in\mathbb{R}^{2N_b}$ represent the weight and bias of the dynamic linear layer, and $\mathbf W^\text{a}_j\in\mathbb{R}^{2N_b\times N_z}$ and $\mathbf b^\text{a}_j\in\mathbb{R}^{2N_b}$ represent the instance-dependent weight and bias adjustments for the parameters of the dynamic linear layer, which are generated by the weight-adjusted MLP and bias-adjusted MLP, respectively. When the feature representation is mapped from the latent space to the label space, the parameters of the dynamic linear layer depend on the input CSI sequence, enabling LPCNet to model the distribution discrepancies between the input CSI sequences caused by the non-stationarity of wireless channels.
	
	\subsubsection{Parameter Adjustment Module}
	
	Generally, the parameters of the prediction branch are all fixed once effectively trained offline. To adapt to the changing channel environment, the online training strategy was utilized in \cite{drnn}. However, the requirement of the online data collection and the huge complexity overhead caused by frequently retraining puts a heavy burden. Therefore, designing low-complexity hypernetworks to dynamically and automatically adjust the NN parameters based on the characteristics of each input CSI sequence is a promising direction \cite{dynn}.
	
	The hypernetwork branch of LPCNet contains two parallel parts for adjusting the weight and bias of the dynamic linear layer, respectively, as shown in Fig. \ref{Fig3}. The architecture of both parts is the MLP network with a single hidden layer, thanks to its universal approximation capability \cite{uab}. The number of neurons of the input layer, hidden layer and output layer of the weight-adjusted MLP are set as $N_i$, $N_{w}$ and $N_z$, respectively. Mathematically, the $k$-th row of the instance-dependent weight matrix $\mathbf W^\text{a}_j$ is derived by 
	\begin{equation}\label{wamlp}
	\mathbf W^\text{a}_j[k,:]=\mathbf W_2^\text{w}\sigma_\text{r}(\mathbf W_1^\text{w} \mathbf x_k+\mathbf b_1^\text{w})+\mathbf b_2^\text{w},
	\end{equation}
	where $\mathbf x_k=[\tilde{\mathbf{h}}_{j,1}[k],\tilde{\mathbf{h}}_{j,2}[k],\ldots,\tilde{\mathbf{h}}_{j,K-1}[k]]^T\in\mathbb{R}^{{K-1}}$ is the input to the weight-adjusted MLP, $\mathbf W_2^\text{w}\in\mathbb{R}^{N_z\times N_w}$ and $\mathbf W_1^\text{w}\in\mathbb{R}^{N_w\times N_i}$ are the weighs of the first and second layers, $\mathbf b_2^\text{w}\in\mathbb{R}^{N_z}$ and $\mathbf b_1^\text{w}\in\mathbb{R}^{N_w}$ are the corresponding biases, while $\sigma_\text{r}(\cdot)$ is the rectified linear unit (ReLU) activation function. Obviously, $N_i$ equals to $K-1$. For $k=1,\ldots,2N_b$, all the parameters of the weight-adjusted MLP are shared to reduce the total number of parameters. By considering that the $k$-th row of the weight $\mathbf W$ of the dynamic linear layer is only used to acquire the $k$-th element of the predicted CSI vector, i.e., $\hat{\bar{\mathbf{h}}}_{j,K+1}[k]$, we thus exploit the historical information of the $k$-th element to generate the instance-dependent weight adjustment $\mathbf W^\text{a}_j[k,:]$, resulting in a low complexity of the weight-adjusted MLP. Similar to the idea of the weight-adjusted MLP, the bias-adjusted MLP also shares all parameters when producing the bias adjustment $\mathbf b^\text{a}_j$. Denote the number of neurons of the hidden layer of the bias-adjusted MLP as $N_{s}$, its mathematical expression can be written as
	\begin{equation}\label{bamlp}
	\mathbf b^\text{a}_j[k]=\mathbf w_2^\text{b}\sigma_\text{r}(\mathbf W_1^\text{b} \mathbf x_k+\mathbf b_1^\text{b})+b_2^\text{b},
	\end{equation}
	where $\mathbf w_2^\text{b}\in\mathbb{R}^{1\times N_s}$, $b_2^\text{b}\in\mathbb{R}$, $\mathbf W_1^\text{b}\in\mathbb{R}^{N_s\times N_i}$ and $\mathbf b_1^\text{b}\in\mathbb{R}^{N_s}$ are the parameters of different layers. With the help of the weight-adjusted MLP and bias-adjusted MLP, the partial parameters of the prediction branch become dynamic and the parameter space is consequently enlarged, facilitating accurate channel prediction in modern wireless communication networks.
	
	\subsubsection{Training of LPCNet}
	
	The prediction and hypernetwork branches of LPCNet are jointly trained in an end-to-end manner. Moreover, the normalized MSE (NMSE) loss function is selected to optimize LPCNet during the training procedure, which is expressed as 
	\begin{equation}\label{loss}
	L(\Psi)=\frac{1}{\mathcal{I}}\sum_{i=1}^{\mathcal{I}}\frac{\Vert \hat{\bar{\mathbf{h}}}_i - \bar{\mathbf{h}}_i \Vert_2^2}{\Vert \bar{\mathbf{h}}_i \Vert_2^2},
	\end{equation}
	where the subscript $i$ denotes the $i$-th sample in the training data, $\mathcal{I}$ is the volume of the training data, while $\Psi$ denotes the parameter set of LPCNet, which is randomly initialized to be $\Psi_0$. The adaptive moment estimation (ADAM) algorithm is applied as the parameter optimizer. The hidden state $\mathbf z_{j,0}$ and cell state $\mathbf c_{j,0}$ are initialized as the zero vector. 
	
	\subsubsection{Complexity Analysis}
	
	The number of parameters of LPCNet comes from two parts: the prediction branch and the hypernetwork branch. In the prediction branch, the number of parameters in LSTM is $4(2N_bN_z+N_z^2+N_z)$ and $2(N_z+1)N_b$ for the dynamic linear layer. In the hypernetwork branch, the weight-adjusted MLP and bias-adjusted MLP have $N_iN_w+N_wN_z+N_w+N_z$ and $N_iN_s+2N_s+1$ parameters, respectively. Therefore, the total number of parameters of LPCNet is $10N_bN_z+4N_z^2+N_iN_s+N_iN_w+N_wN_z+5N_z+2N_b+2N_s+N_w+1$. The parameter sharing scheme in both the weight-adjusted MLP and bias-adjusted MLP reduces significantly the number of parameters in the hypernetwork branch.
	
	We now assess the computational complexity of LPCNet, which mainly stems from the dynamic parameters generation in the parameter adjustment module and the update of the latent states in LSTM. The computational complexity of the parameter adjustment module is $\mathcal{O}(N_b(N_iN_w+N_wN_z+N_iN_s))$, while the computational complexity of LSTM is $\mathcal{O}(N_i(N_bN_z+N_z^2))$. In total, the computational complexity of LPCNet is $\mathcal{O}(N_iN_z^2+N_b(N_iN_z+N_iN_w+N_wN_z+N_iN_s))$.
	
	Given the increasing number of antennas in 5G and future 6G networks, the growing number of BS antennas makes $N_b$ non-negligible, further leading to an increase in $N_z$ in the subsequently feature extraction part. By contrast, the very limited input length $N_i$ allows for relatively small values of $N_w$ and $N_s$. The impact of different hidden sizes on the CSI prediction accuracy will be compared and analyzed in Section \uppercase\expandafter{\romannumeral5}-C. In summary, it is concluded that the complexity of LPCNet is mainly dominated by $N_b$ and $N_z$.
	
	\subsection{JLPCNet}
	
	As described in problem (\ref{dynamicnn}), the purpose of channel prediction is to acquire the entire CSI in the future, containing both the amplitude and direction information. However, the design of the BF vector in (\ref{rate}) only focuses on the direction information of wireless channels. As a result, although the CSI prediction error is lower, the achievable downlink  SE may perform worse. With the objective of maximizing the BF performance gain rather than the CSI prediction accuracy, jointly designing channel prediction and BF blocks is a meaningful approach. 
	
	From a perspective of the end-to-end design, joint channel prediction and BF design can be realized by constructing a function map to directly model the relationship between the past CSI sequence and the future BF vector. Taking the past CSI sequence as the input, we can directly predict the future BF vector used for the data transmission phase without explicitly obtaining the future CSI. Besides, the non-stationary property of CSI also needs to be taken into account. We, thus, follow the NN architecture of LPCNet to achieve the joint channel prediction and BF design. The NN input is consistent with LPCNet, i.e., the past CSI sequence $\{\bar{\mathbf{h}}_{j,t}\}_{t=1}^{K}$, but the NN output is the predicted BF vector $\hat{\mathbf{w}}_{j,t}$ instead of the predicted CSI. We refer to this DL based joint channel prediction and BF method as JLPCNet. 
	
	Different from LPCNet focusing on the full channel knowledge, the major objective of JLPCNet is to predict the BF vector with specific direction, and thus the cosine similarity is adopted to design the loss function for end-to-end training JLPCNet. In particular, the cosine similarity $\rho$ is written as  
	\begin{equation}\label{cos}
	\rho = \frac{|\hat{\mathbf{w}}_{j,t}^H \mathbf{w}_{j,t}|}{\Vert\hat{\mathbf{w}}_{j,t}\Vert_2 \Vert \mathbf{w}_{j,t} \Vert_2}.
	\end{equation}
	It is clear that the range of $\rho$ is $[0,1]$. The closer the value of $\rho$ is to 1, the better BF performance can be achieved. Accordingly, the negative of the cosine similarity in (\ref{cos}) is used as the loss function, which can be expressed as 
	\begin{equation}\label{LPBloss}
	\tilde{L}(\tilde{\Psi}) =- \frac{1}{\mathcal{\tilde{I}}}\sum_{i=1}^{\mathcal{\tilde{I}}}
	\frac{|\hat{\mathbf{w}}_{i}^H \mathbf{w}_{i}|}{\Vert\hat{\mathbf{w}}_{i}\Vert_2 \Vert \mathbf{w}_{i} \Vert_2},
	\end{equation}
	where the subscript $i$ denotes the $i$-th sample in the training data, $\mathcal{\tilde{I}}$ denotes the size of the training data, $\tilde{\Psi}$ denotes the parameter set of JLPCNet. The choice of the parameter initialization and the optimizer of JLPCNet is the same as LPCNet. 
	
	\section{Simulation Results and Analysis}
	
	In this section, we first elaborate on the simulation settings, including the channel data generation, hyperparameter settings, baseline and evaluation metric. Then, we demonstrate the effectiveness and flexibility of the proposed framework. Finally, we evaluate the proposed methods, i.e., LPCNet and JLPCNet, in different scenarios for fixed and varying UE speeds. 
	
	\subsection{Simulation Settings}
	
	\subsubsection{Channel Data Generation}
	
	QuaDRiGa channel generator \cite{Quadriga} is widespread in simulating time-varying channels and can be viewed as a 3GPP 38.901 reference implementation, which follows the geometry-based stochastic channel modeling (GBSM) approach. Therefore, we employ QuaDRiGa for numerical simulation. Both the UMA-LOS and UMA-NLOS scenarios are considered for the performance evaluation of the proposed LPCNet and JLPCNet. For the above used two scenarios, the number of paths $L$ is 12 and 21, respectively. The communication frequency $f$ is set as 3.5GHz and the height of the BS is set as 25m. The number of antennas in both rows and columns of the BS is set as 4, i.e., $N_l=4$ and $N_r=4$, while the polarization angles are set as $\pm45^{\circ}$. The UE is assumed to move along a linear trajectory with a fixed speed $v$, and the period of the sounding reference signal (SRS) is set as 2ms. The scattering environment is reconstructed 60 times and the way of sliding window is adopted to build the channel dataset. We split the generated dataset into two parts, namely, the training and test datasets. The sizes of the training dataset and the test dataset are 110,000 and 25,000, respectively. The NN parameters are optimized based on the training dataset and the online performance is evaluated on the test dataset. For clarity, the basic simulation parameters are summarized in Table \ref{sp}.
	
	\begin{table}[t]
		\small
		\centering
		\caption{Basic simulation parameters.}\label{sp}
		\renewcommand\arraystretch{1.3}
		\begin{tabular}{c|c}
			\hline
			Parameter & Value \\
			\hline\hline
			Channel generator & QuaDRiGa [43] \\
			\hline
			\multirow{2}{*}{Communication scenarios} & UMA-LOS  \\
			\cline{2-2}
			& UMA-NLOS \\
			\hline
			Communication frequency, $f$ &  3.5GHz \\
			\hline
			Number of antennas in a column, $N_l$ & 4 \\
			\hline
			Number of antennas in a row, $N_r$ & 4 \\
			\hline
			Polarization angles & $\pm45^{\circ}$ \\
			\hline
			BS height & 25m \\
			\hline
			UE trajectory & Linear track  \\
			\hline
			Period of SRS & 2ms \\
			\hline 
		\end{tabular}
	\end{table}
	
	\begin{table}[t]
		\small
		\centering
		\caption{Hyperparameter settings of LPCNet.}\label{hs}
		\renewcommand\arraystretch{1.3}
		\begin{tabular}{c|c}
			\hline
			Parameter & Value \\
			\hline\hline
			Input length, $K$ & 15 \\
			\hline
			Number of neurons of LSTM, $N_z$ & 256 \\
			\hline
			Number of neurons of the weight-adjusted MLP, $N_s$ & 64 \\
			\hline 
			Number of neurons of the bias-adjusted MLP,  $N_w$ & 64 \\
			\hline
			Learning rate & 0.0001 \\
			\hline
			Batch size & 200 \\
			\hline
			Epochs & 1000 \\
			\hline
		\end{tabular}
	\end{table}
	
	\subsubsection{Hyperparameter Settings}
	
	The length $K$ of the input CSI sequence used for prediction is set to 15. For the prediction NN branch, the number of neurons $N_z$ of LSTM is set to 256. For the hypernetwork branch, the hidden sizes $N_w$ and $N_s$ of the weight-adjusted MLP and bias-adjused MLP are both set to 64. During the training stage, the ADAM optimizer with the default hyperparameter setting is used to update the parameters. The learning rate, number of epochs and batch size are set to 0.0001, 1,000 and 200, respectively. The hyperparameter settings of LPCNet are listed in Table \ref{hs}. All experiments are conducted on PyTorch. 
	
	\subsubsection{Baselines and Evaluation Metric}
	
	Our reference channel prediction methods are the sample-and-hold (SH) approach, the AR predictor \cite{AR} and the LSTM based method \cite{lstm}. The SH approach treats the latest CSI vector as the prediction result. The AR order is set to 5 and the length of the past CSI sequence for its parameter calculation is set to 15. The AR parameters are updated for every window. The hyperparameter setting of the LSTM based method is consistent with LPCNet. Based on the predicted CSI, the zero-forcing (ZF) algorithm is used to compute the BF vector. NMSE is utilized to evaluate the CSI prediction accuracy, and the cosine similarity is considered as the evaluation metric for the BF performance.
	
	Unless stated otherwise, the simulation settings mentioned above are applied throughout this section.
	
	\subsection{Evaluation for the Proposed Framework}
	
	\begin{figure}[t]
		\centerline{\includegraphics[width=3.5in]{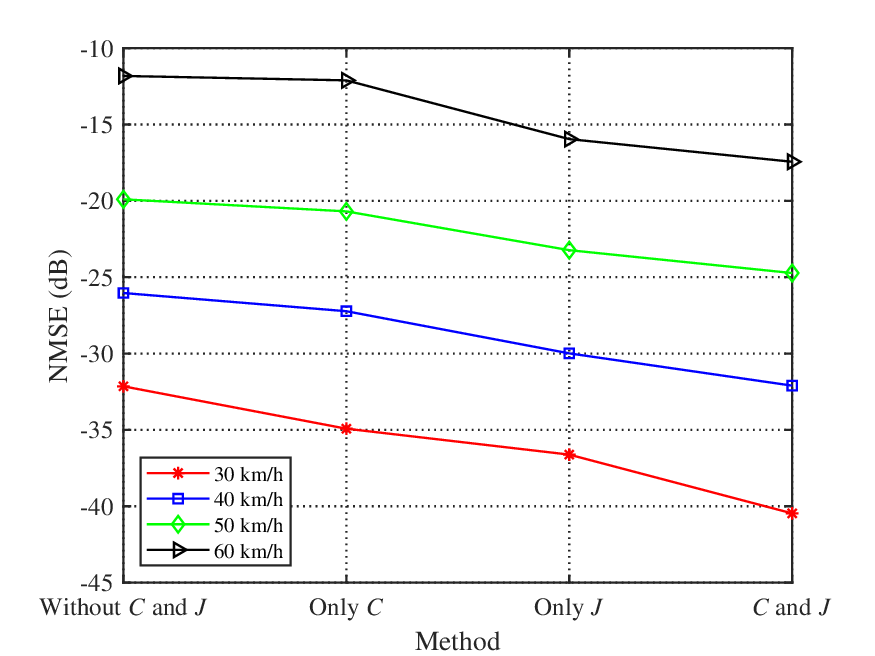}}
		\caption{Ablation on the preprocessor $C$ and adjuster $J$. Only $C$ means to employ the difference operator in (\ref{diff}) to mitigate the non-stationarity of the input CSI sequence; only $J$ means to employ the hypernetwork branch to model the distribution discrepancies between the input CSI sequences; $C$ and $J$ mean to employ both, and without $C$ and $J$ mean to not explicitly capture the CSI non-stationarity. The communication scenario is UMA-NLOS and the prediction length is 2ms.}
		\label{Fig6}
	\end{figure}
	
	\subsubsection{Effects of the Preprocessor and Adjuster Modules}
	
	To investigate the effect of modeling the CSI non-stationarity with the adjuster $J$ and preprocessor $C$, we compare the NMSE performance obtained by these LPCNet variants: without $C$ and $J$, only employing $C$, only employing $J$ and employing both. The experiment is conducted for different UE speeds. As shown in Fig. \ref{Fig6}, whether employing $C$ or $J$ would be helpful for the improvement of the CSI prediction accuracy. More significant performance gains can be observed by using $J$ than $C$, which demonstrates the necessity to account for the temporal distribution shift problem. Moreover, the best CSI prediction accuracy is achieved if we use both $C$ and $J$, which validates the effectiveness of our framework.
	
	\subsubsection{Flexibility Analysis}
	
	\begin{figure}[t]
		\centerline{\includegraphics[width=3.5in]{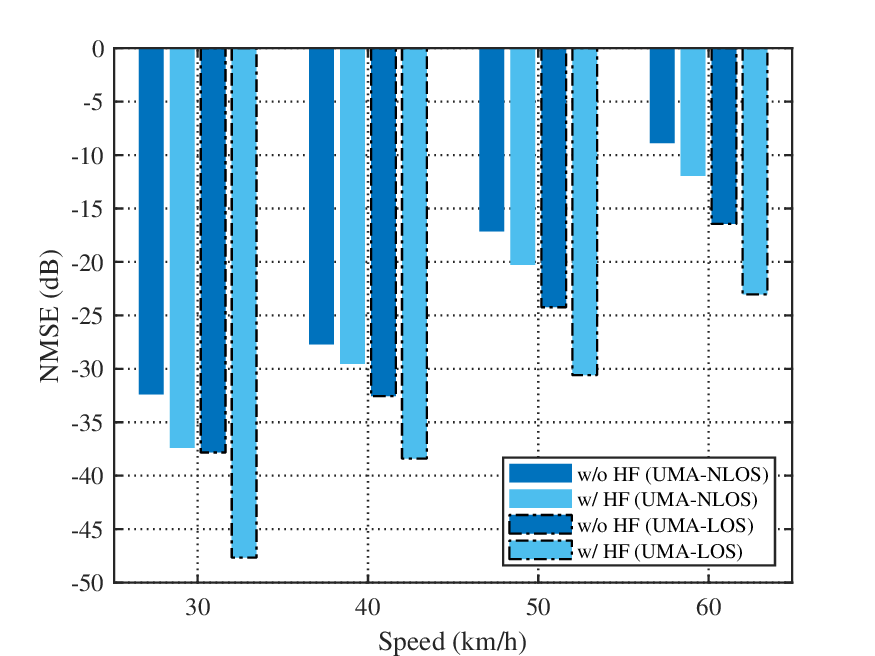}}
		\caption{Flexibility analysis. Method w/o HF denotes the basic self-attention architecture without our proposed framework and method w/ HF denotes that the self-attention is used as the encoder $E$ in our proposed framework. The face color is employed for method discrimination, while the edge line is employed to distinguish between the communication scenarios. The prediction length is 2ms.}
		\label{Fig9}
	\end{figure}
	
	In this subsection, we explore the flexibility of the proposed hypernetwork based framework (HF) for non-stationary channel prediction. Apart from the LSTM model used in LPCNet, the self-attention architecture \cite{selfattention}, also known as a popular deep series modeling approach, is incorporated into the proposed framework for accurate channel prediction. Specifically, the self-attention architecture is used as the encoder $E$ of our framework for the modeling of temporal dependencies, and the rest modules are consistent with LPCNet. In the encoder $E$, the number of parallel attention heads is set as 4, and the dimensions of queries, keys and values in the scaled dot-product attention are both set as 32. The inner-layer in the position-wise feed-forward network has 128 neurons. As shown in Fig. \ref{Fig9}, the pilot study verifies that the CSI prediction accuracy of the self-attention architecture can be significantly boosted by incorporating it into our proposed framework. On average, w/ HF can obtain performance gains of 7.4dB (at 30km/h), 3.8dB (at 40km/h), 4.7dB (at 50km/h) and 4.8dB (at 60km/h), respectively.
	
	\subsection{Performance of LPCNet}
	
	\subsubsection{Complexity Comparison}
	
	\begin{figure}[t]
		\centerline{\includegraphics[width=3.8in]{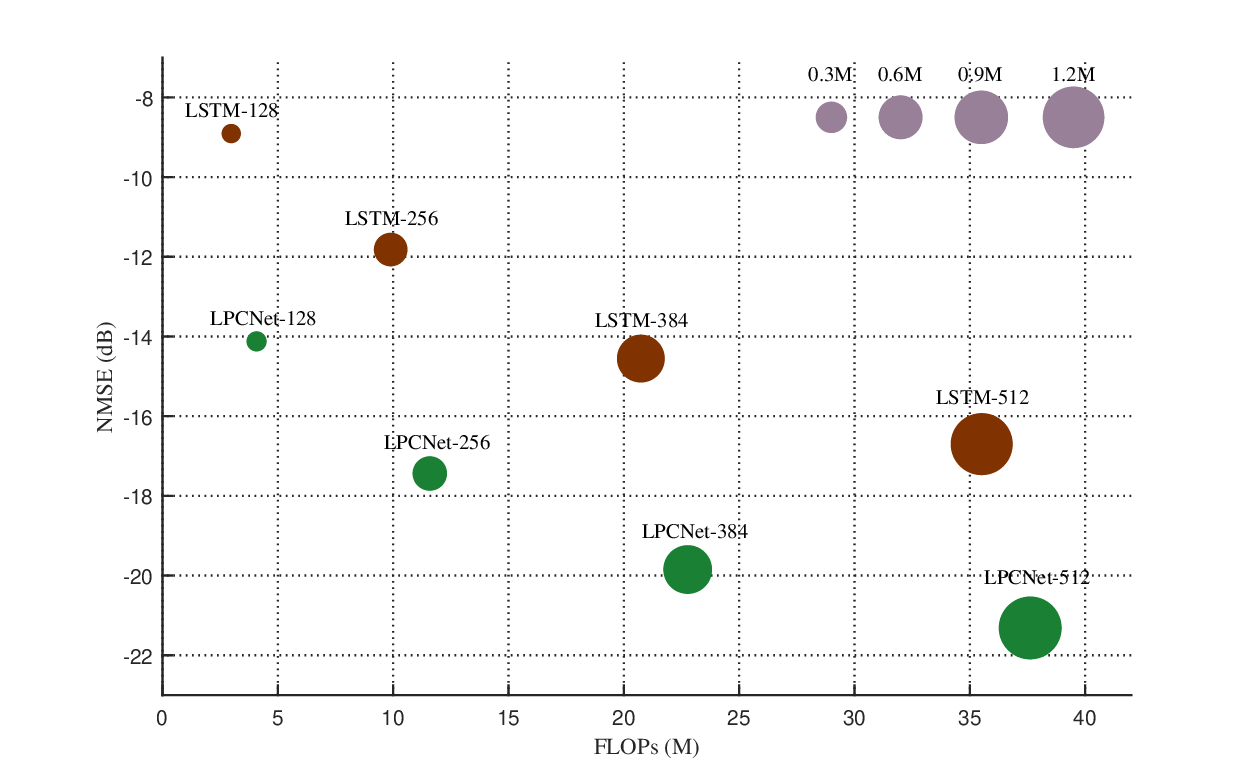}}
		\caption{NMSE performance versus number of FLOPs of LSTM and LPCNet with different hidden sizes $N_z$, which vary in $\{128,256,384,512\}$. The prediction length is 2ms, the UE speed is 60km/h, and the communication scenario is UMA-NLOS. The size of circles represents the number of parameters of NNs.}
		\label{FigR1Com}
	\end{figure}
	
	The complexity of NNs is measured by the number of parameters and floating point operations (FLOPs), respectively. The comparison results between LSTM and LPCNet with different hidden sizes $N_z$ are shown in Fig. \ref{FigR1Com}. Thanks to the parameter sharing in LPCNet, it can be found that LPCNet increases the parameters space only moderately, when its hidden size $N_z$ is the same as LSTM. In terms of the computational complexity, for any $N_z$, the increase of FLOPs brought by LPCNet is very limited as well. More importantly, LSTM-512 achieves a comparable NMSE performance with LPCNet-256, but the number of FLOPs of LSTM-512 is more than three times that of LPCNet-256, as well as the number of parameters.
	
	\subsubsection{Hyperparameter Sensitivity}
	
	\begin{table}[t]
		\centering
		\caption{The NMSE (dB) performance of LPCNet under different choices of the number of neurons $N_w$ in the weight-adjusted MLP. The result is based on the UMA-NLOS scenario and a 2ms prediction length.}\label{hp}
		\renewcommand\arraystretch{1.3}
		\begin{tabular}{c|c|c|c|c}
			\hline
			Number of neurons & 30km/h& 40km/h& 50km/h& 60km/h\\
			\hline\hline
			$N_w=32$& -40.4601& -30.7737& -24.6199& -17.3912\\
			\hline
			$N_w=64$& \textbf{-40.4685}& -32.1025& -24.7276& -17.4392\\
			\hline
			$N_w=128$& -39.8539& \textbf{-32.8023}& \textbf{-25.2927}& \textbf{-17.5049}\\
			\hline
		\end{tabular}
	\end{table}
	
	In LPCNet, one of the core architecture hyperparameters is the number of neurons $N_w$ in the weight-adjusted MLP. We thus analyze the robustness of LPCNet against the number of neurons $N_w$, which varies as $\{32,64,128\}$. The results are listed in Table \ref{hp}. It is seen that by increasing the number of neurons $N_w$, the CSI prediction accuracy of LPCNet is improved when the UE speed varies in $\{40,50,60\}$km/h. Therefore, when deploying LPCNet in high mobility scenarios, the large number of neurons $N_w$ would be a better choice to achieve a high CSI prediction accuracy. 
	
	\subsubsection{Results in the Fixed Speed Scenarios}
	
	\begin{figure*}[t]
		\centering
		\subfloat[UMA-LOS.]{\includegraphics[width=3.5in]{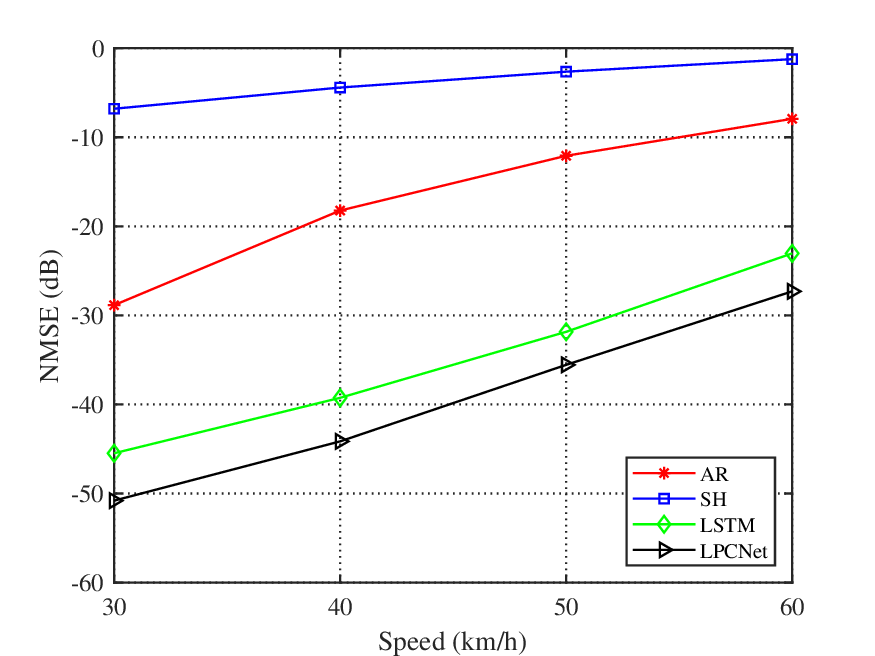}%
			\label{fig4a}}
		\hfil
		\subfloat[UMA-NLOS.]{\includegraphics[width=3.5in]{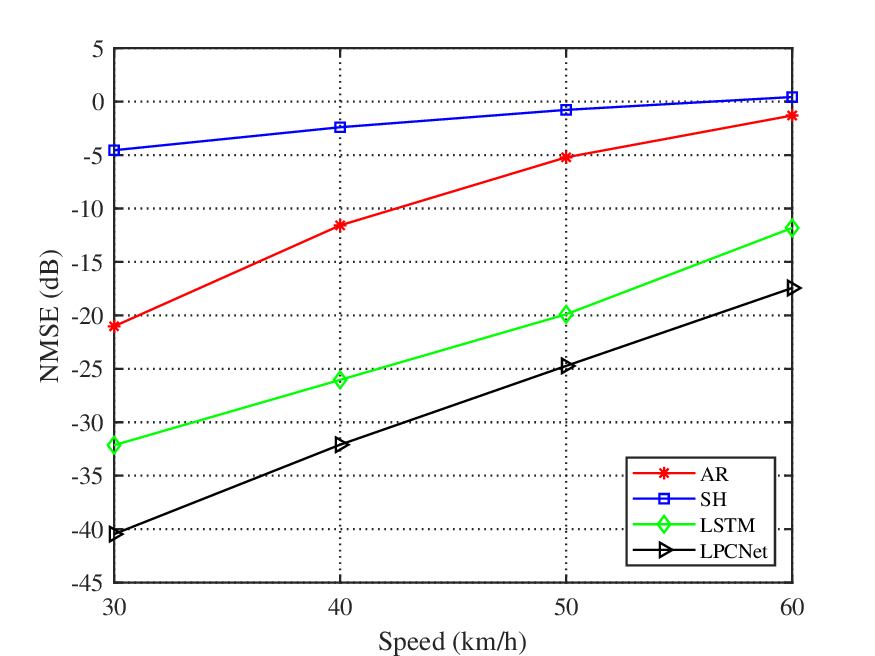}%
			\label{fig4b}}
		\caption{NMSE (dB) vs speed (km/h) for two scenarios when the prediction length is 2ms.}
		\label{fig4}
	\end{figure*}
	
	\begin{figure*}[t]
		\centering
		\subfloat[UMA-LOS.]{\includegraphics[width=3.5in]{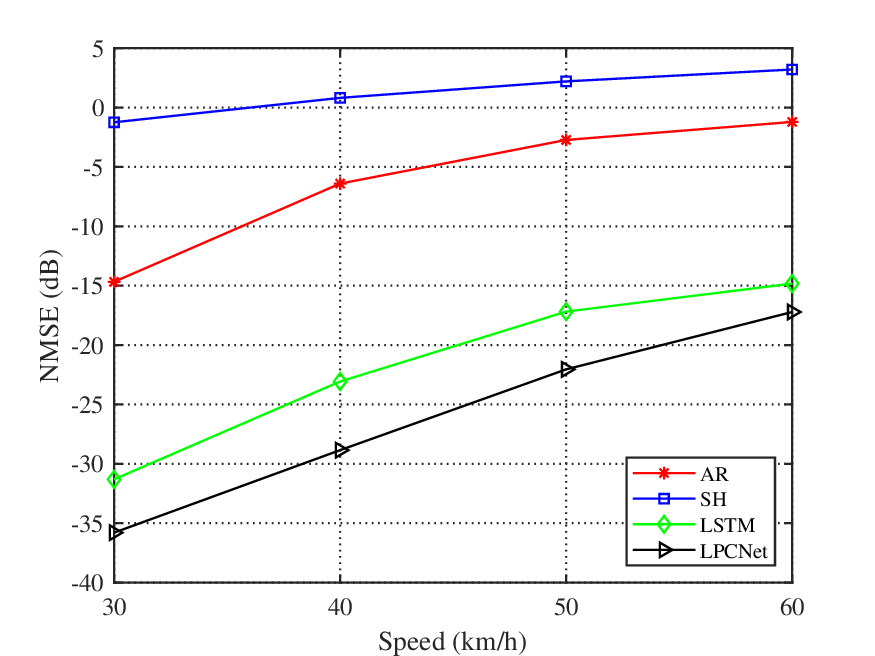}%
			\label{fig5a}}
		\hfil
		\subfloat[UMA-NLOS.]{\includegraphics[width=3.5in]{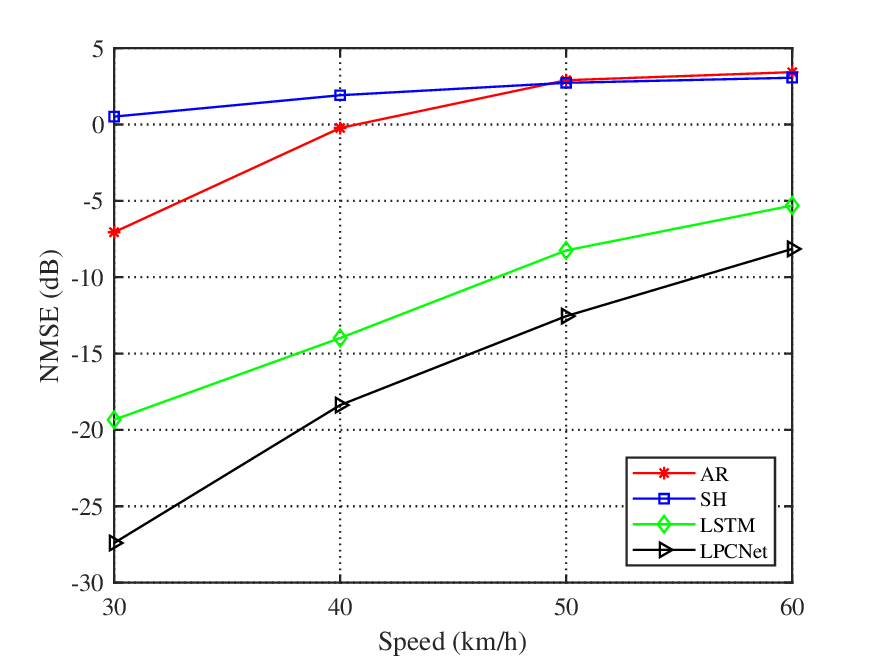}%
			\label{fig5b}}
		\caption{NMSE (dB) vs speed (km/h) for two scenarios when the prediction length is 4ms.}
		\label{fig5}
	\end{figure*}
	
	Figure \ref{fig4} shows the NMSE performance of all channel prediction methods with respect to the speed of the UE for two different scenarios. From both Fig. \ref{fig4a} and Fig. \ref{fig4b}, it is seen that the NMSE performance of the algorithms declines with the increase of the UE speed. Compared to any baseline, LPCNet has significant performance gains for all considered speeds. In the UMA-NLOS scenario, our LPCNet achieves better results than the LSTM, AR and SH  by reducing the NMSE $22.8\%$, $70.9\%$ and $95.2\%$ on average. The reason is that LPCNet not only has a powerful nonlinear learning capability, but also considers the CSI non-stationarity. Both LSTM and AR perform better than SH, and LSTM outperforms AR obviously. The poor performance of SH demonstrates that the wireless channel changes rapidly over time.
	
	Furthermore, we now increase the prediction length from 2ms to 4ms and the comparison results are illustrated in Fig. \ref{fig5}. It can be observed that LPCNet still performs pretty well in this case. Compared to LSTM, LPCNet achieves at least 2.4dB and 2.8dB performance improvement in the UMA-LOS and UMA-NLOS scenarios. For the reference methods, LSTM exhibits superior performance over AR and SH. In the UMA-LOS scenario, AR is able to perform better than SH. However, in the UMA-NLOS scenario, the performance of AR degrades severely, even worse than SH at the high UE speed, which indicates the limited prediction capability of the AR model.
	
	\subsubsection{Results in the Varying Speed Scenarios}
	
	\begin{figure*}[t]
		\centering
		\subfloat[UMA-LOS.]{\includegraphics[width=3.5in]{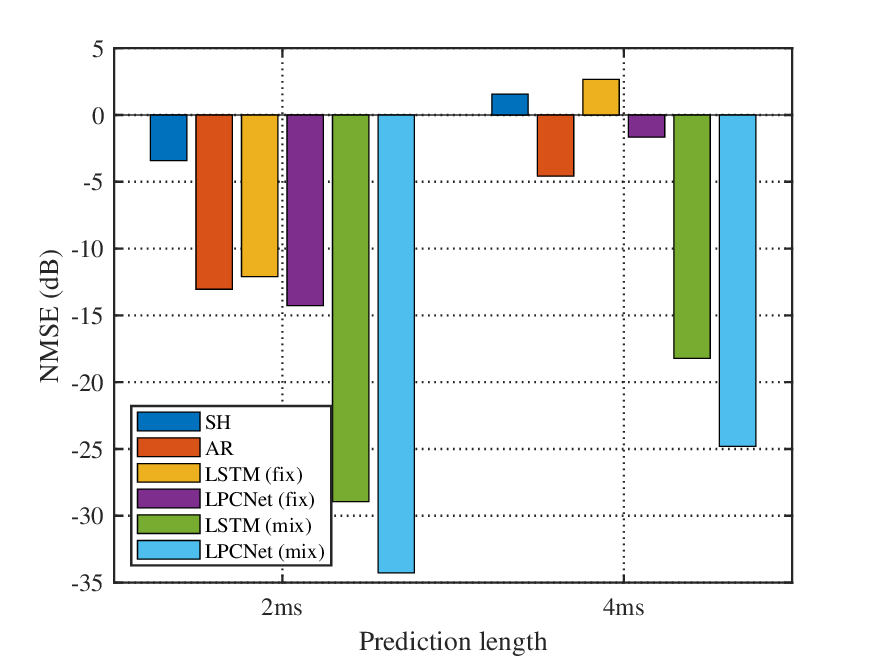}%
			\label{fig7a}}
		\hfil
		\subfloat[UMA-NLOS.]{\includegraphics[width=3.5in]{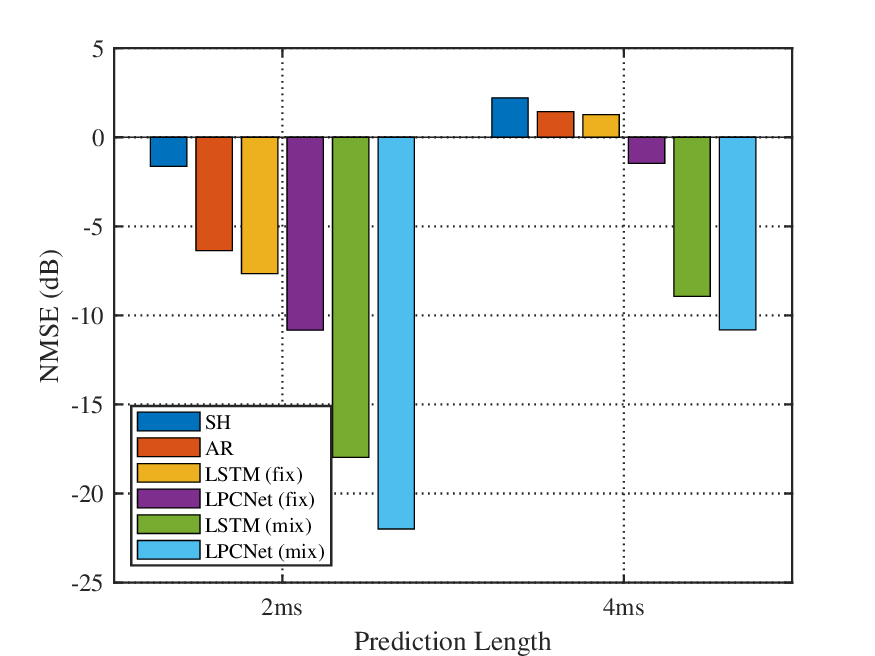}%
			\label{fig7b}}
		\caption{NMSE (dB) vs prediction length (ms) for two scenarios when the UE speed varies in the range of [30,60]km/h.}
		\label{fig7}
	\end{figure*}
	
	In addition to assuming that the UE speed is fixed, we study the CSI prediction accuracy of the algorithms for varying UE speeds. In the process of regenerating scattering environments, the UE speed is increased from 30km/h to 60km/h at equal intervals. In other words, the UE speed becomes higher at every scattering environment reconstruction. Therefore, we can obtain a new pair of the training and test datasets, both including samples of various UE speeds in the range of [30,60]km/h.
	
	The comparison result regarding the varying UE speeds is shown in Fig. \ref{fig7}. LSTM (fix) and LPCNet (fix) are trained on the dataset with the UE speed $v=60$km/h, while LSTM (mix) and LPCNet (mix) are trained on the new dataset with various UE speeds. All channel prediction methods are evaluated on the new test dataset. From both Fig. \ref{fig7a} and \ref{fig7b}, it is not surprising that LPCNet (mix) shows a superior performance than its counterparts for all scenarios and prediction lengths. Moreover, LPCNet (fix) performs better than LSTM (fix), once again proving the learning capability of LPCNet. Besides, both LPCNet (fix) and LSTM (fix) perform worse than AR in the setting of the UMA-LOS scenario and 4ms prediction length, which indicates that the richness of the training dataset is also critical for DL based methods.
	
	\begin{figure*}[t]
		\centering
		\subfloat[UMA-LOS.]{\includegraphics[width=3.5in]{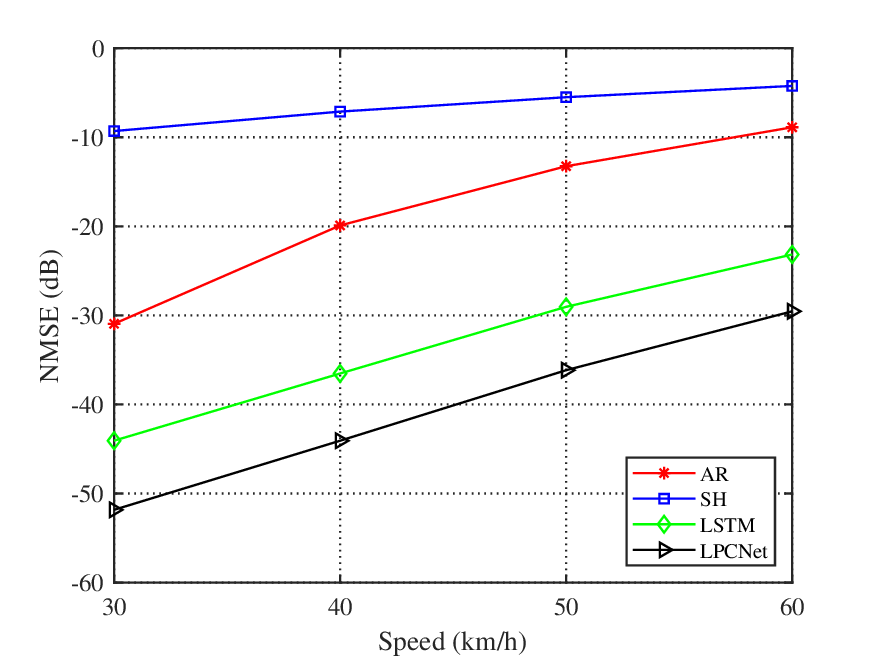}%
			\label{R1a}}
		\hfil
		\subfloat[UMA-NLOS.]{\includegraphics[width=3.5in]{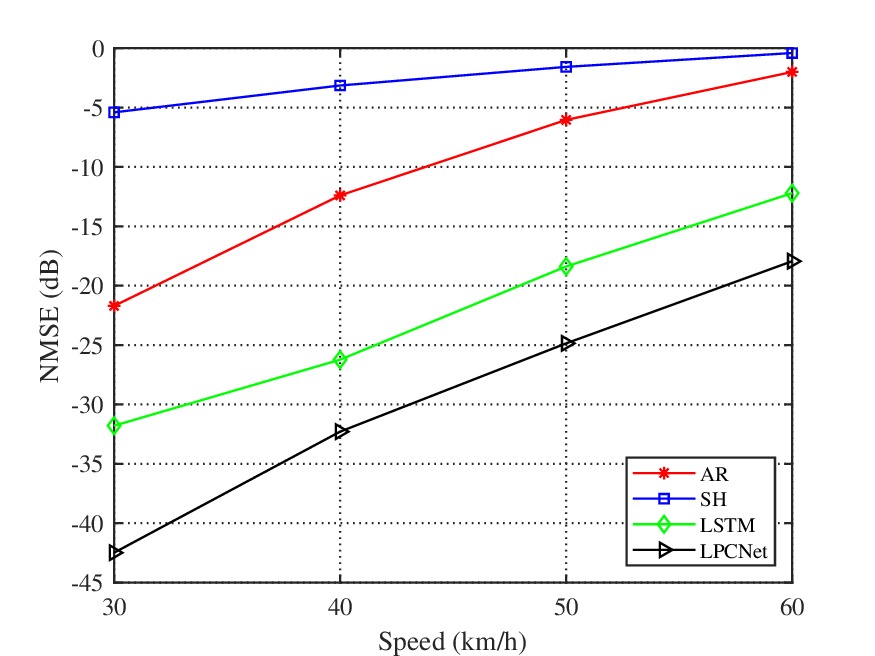}%
			\label{R1b}}
		\caption{NMSE (dB) vs speed (km/h) for the circular UE trajectory when the prediction length is 2ms.}
		\label{figR1}
	\end{figure*}
	
	\subsubsection{Results in other types of UE trajectories}Except for the linear UE trajectory, we here additionally consider that the UE moves along a circular trajectory, and the remaining simulation settings are the same with the linear one. Compared to the linear track, the circular track means that the travel azimuth angle in (\ref{v}) is time-varying. Figure \ref{figR1} shows the NMSE performance of LPCNet against the compared predictors in case of the circular UE trajectory. From the figure, it is clearly seen that LPCNet performs significantly better than the rest of predictors. The results demonstrate that our LPCNet is not only applicable to linear trajectories, but also to nonlinear trajectories. Among all baselines, LSTM consistently outperforms AR, while SH ranks last.
	
	\subsection{Performance of JLPCNet}
	
	\begin{table*}[t]
		\centering
		\caption{Comparison of the BF performance between JLPCNet and other algorithms. The cosine similarity is used as the evaluation metric.}\label{JLPCNet}
		\renewcommand\arraystretch{1.3}
		\begin{tabular}{c|c|c|c|c|c}
			\hline
			\multirow{2}{*}{Speed (km/h)}& \multirow{2}{*}{Method}& \multicolumn{2}{c|}{UMA-LOS}& \multicolumn{2}{c}{UMA-NLOS}\\
			\cline{3-6}
			&& 2ms &4ms &2ms & 4ms\\
			\hline\hline
			\multirow{5}{*}{60}
			&SH+ZF&85.4004$\%$&71.5193$\%$&73.4645$\%$&59.9736$\%$\\
			\cline{2-6}
			&AR+ZF&93.3715$\%$&74.6731$\%$&76.7773$\%$&50.7871$\%$\\
			\cline{2-6}
			&LSTM+ZF&99.7779$\%$&98.4355$\%$&97.5748$\%$&86.3250$\%$\\
			\cline{2-6}
			&LPCNet+ZF&99.9138$\%$&99.1179$\%$&99.2991$\%$&93.6901$\%$\\
			\cline{2-6}
			&JLPCNet&\textbf{99.9280$\%$}&\textbf{99.4171$\%$}&\textbf{99.3461$\%$}&\textbf{93.8877}$\%$\\
			\hline
			\multirow{5}{*}{[30,60]}
			&SH+ZF&90.8691$\%$&78.6081$\%$&82.0028$\%$&65.1793$\%$\\
			\cline{2-6}
			&AR+ZF&97.9003$\%$&87.8044$\%$&92.5904$\%$&70.7276$\%$\\
			\cline{2-6}
			&LSTM+ZF&99.9450$\%$&99.2814$\%$&99.4223$\%$&94.6938$\%$\\
			\cline{2-6}
			&LPCNet+ZF&99.9836$\%$&99.8459$\%$&99.7816$\%$&96.8316$\%$\\
			\cline{2-6}
			&JLPCNet&\textbf{99.9932}$\%$&\textbf{99.8675$\%$}&\textbf{99.8671$\%$}&\textbf{97.7518$\%$}\\
			\hline
		\end{tabular}
	\end{table*}
	
	In this subsection, we study the BF performance of our joint channel prediction and BF design in terms of the cosine similarity. The considered algorithms treat the channel prediction problem and BF problem as two independent parts, named SH+ZF, AR+ZF, LSTM+ZF, and LPCNet+ZF. Our JLPCNet directly predicts the future BF vector based on the past CSI sequence. 
	
	Table \ref{JLPCNet} illustrates the cosine similarity performance of all methods in UMA-LOS and UMA-NLOS scenarios, in which different prediction lengths and UE speeds are considered. From the table, it can be seen that JLPCNet performs better than the rest of algorithms in both the UMA-LOS and UMA-NLOS scenarios for fixed and varying UE speeds. The reason is that the design of JLPCNet takes the subsequent BF block into account, and focuses on the direction information rather than the full channel knowledge. LSTM+ZF performs worse than LPCNet+ZF, but clearly better than AR+ZF. On average, SH+ZF yields the worst BF performance, which once more emphasizes the great importance of the prediction.
	
	\section{Conclusion}
	
	In this paper, we proposed a hypernetwork based framework to tackle the non-stationary channel prediction problem rooted in modern wireless communication networks. Our framework is composed of two parallel branches, termed as the prediction NN branch and the hypernetwork branch, respectively. The prediction NN branch models the temporal dependencies of the wireless channel to output the future CSI, while the hypernetwork branch adjusts the parameters of the prediction NN branch over time. Furthermore, we introduced a DL based prediction method, i.e., LPCNet. Specifically, a dynamic linear layer was constructed to capture the data distribution shift of the wireless channel. Moreover,  two parameter sharing light MLP networks were designed to update the parameters in the dynamic linear layer. Additionally, considering the subsequent BF block, we presented a DL based joint channel prediction and BF design, i.e., JLPCNet, to maximize the BF performance rather than the CSI prediction accuracy. Our simulation results showcased that the proposed framework is effective and flexible. Moreover, LPCNet performs better than the existing channel prediction methods in terms of NMSE, whilst JLPCNet achieves a superior performance in terms of the cosine similarity. 
	
	\bibliographystyle{IEEEtran}
	\bibliography{IEEEabrv,NewReferences}
	
 	\begin{IEEEbiography}[{\includegraphics[width=1in,height=1.25in,clip,keepaspectratio]{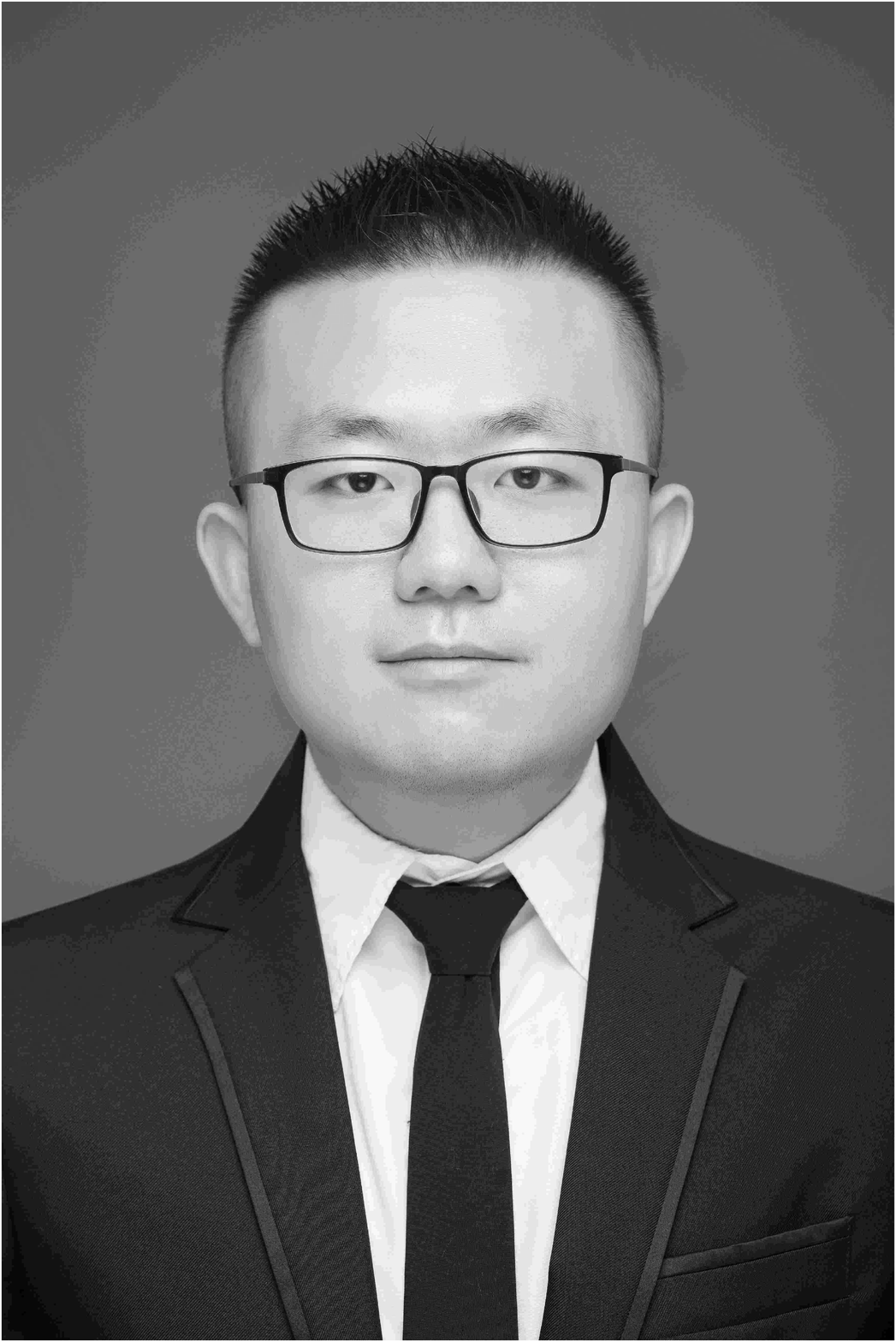}}]{Guanzhang Liu}
 		received the B.S. degree in applied mathematics from Harbin Engineering University, Harbin, China, in 2019. He is currently working toward the Ph.D. degree in applied mathematics with Xi’an Jiaotong University, Xi’an, China. His current research interests include the deep learning for intelligent wireless communications, channel estimation and prediction in high mobility scenarios.
 	\end{IEEEbiography}

 	\begin{IEEEbiography}[{\includegraphics[width=1in,height=1.25in,clip,keepaspectratio]{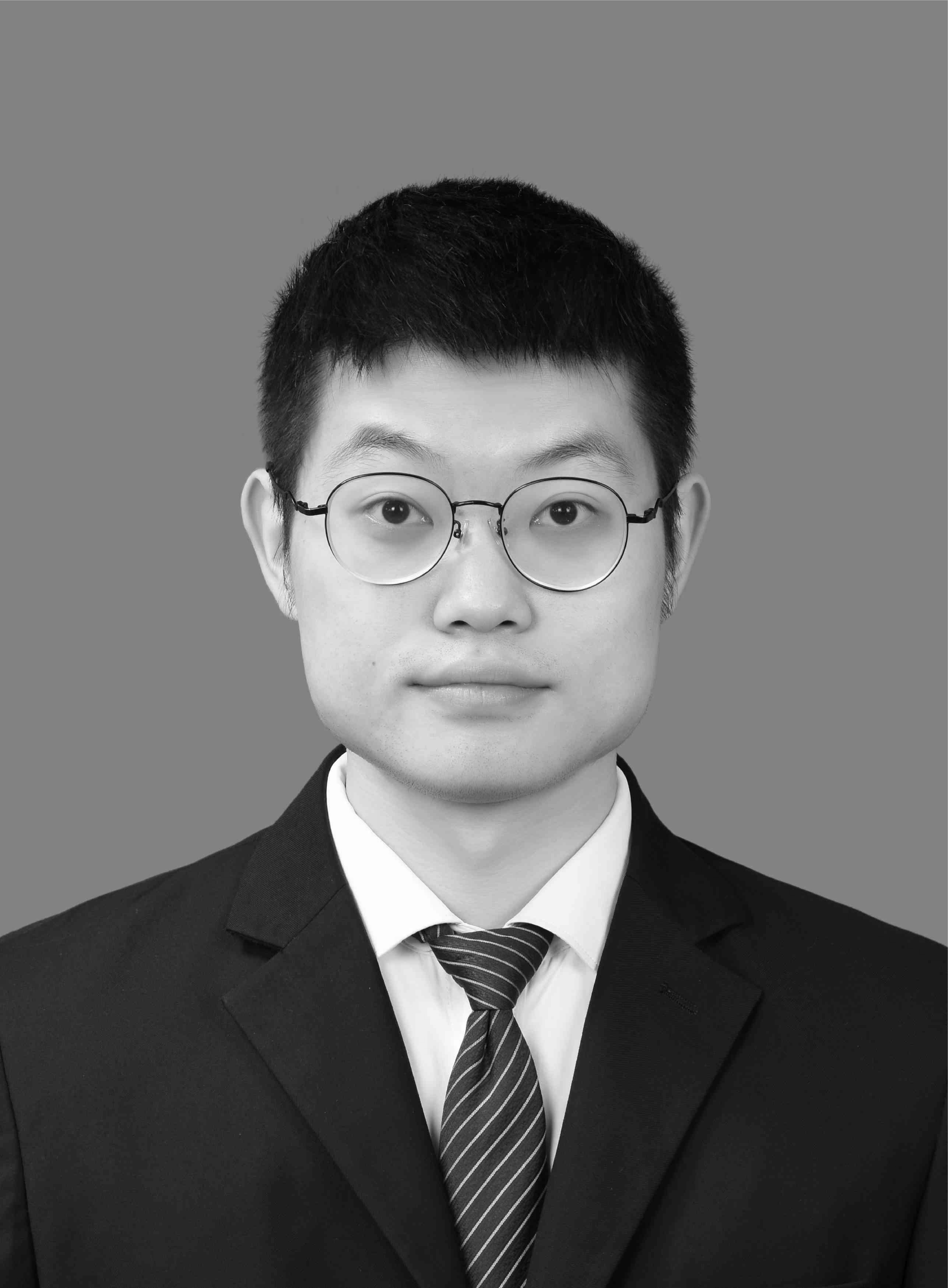}}]{Zhengyang Hu}
 		received the B.S. degree in applied mathematics from Xi'an Jiaotong University, Xi'an, China, in 2019. He is currently working toward the Ph.D. degree in applied mathematics with Xi'an Jiaotong University, Xi'an, China. His current research interests include the deep learning for physical layer, massive MIMO and intelligent wireless communication.
 	\end{IEEEbiography}

 	\begin{IEEEbiography}[{\includegraphics[width=1in,height=1.25in,clip,keepaspectratio]{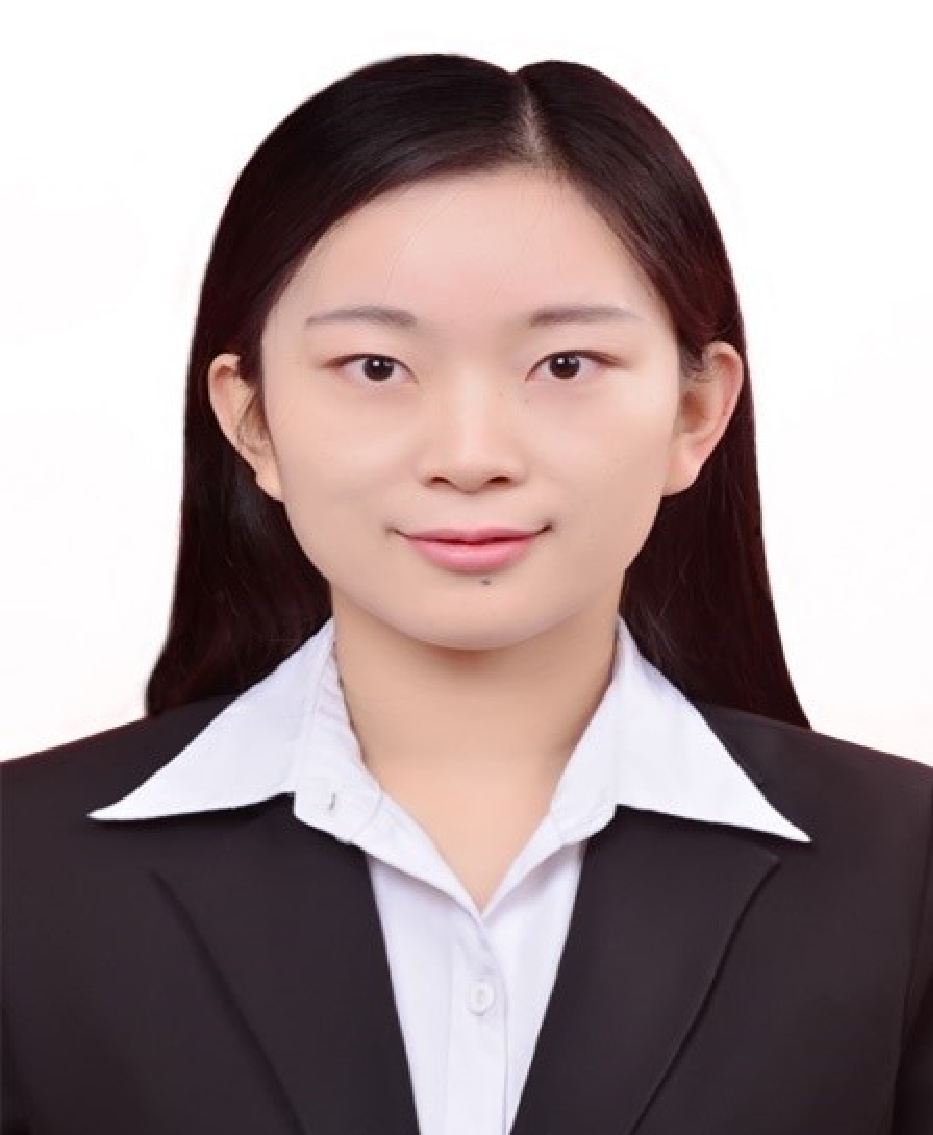}}]{Lei Wang}
 		received the B.S. degree in applied mathematics from Xinjiang University in 2015, the M.S. degree in computational mathematics from Xinjiang University in 2018. She is currently pursuing the Ph.D degree in Mathematics and Statistics at Xi’an Jiaotong University. Her research interests include intelligent wireless communication and machine learning to predict time series problems.
 	\end{IEEEbiography}

 	\begin{IEEEbiography}[{\includegraphics[width=1in,height=1.25in,clip,keepaspectratio]{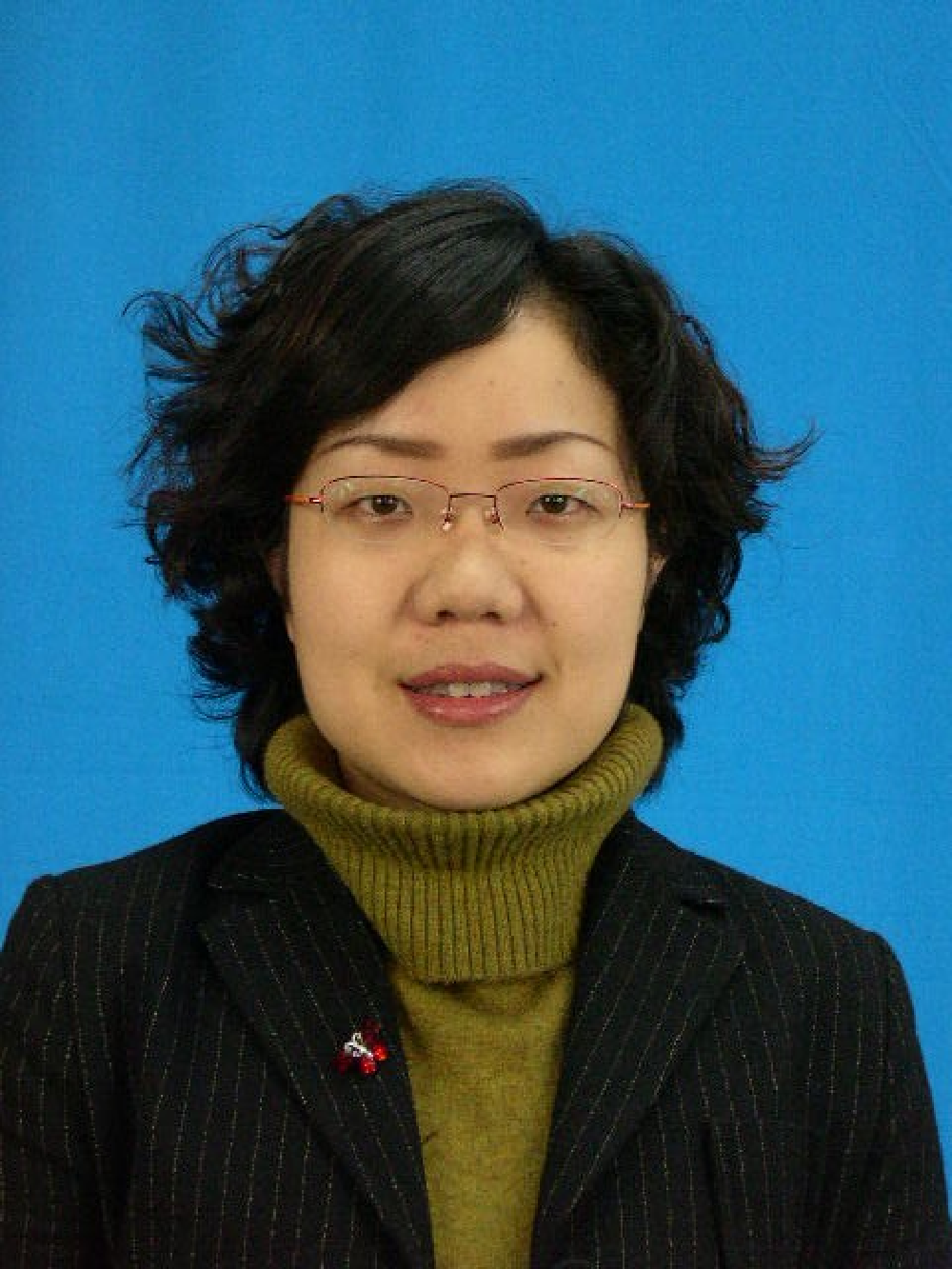}}]{Hongying Zhang}
 		received the M.Sc. and Ph.D. degrees from Xi’an Jiaotong University, Xi’an, China. She is currently a professor with School of Mathematics and Statistics, Xi’an Jiaotong University. Her research interests include artificial intelligence, granular computing, and machine learning.
 	\end{IEEEbiography}

 	\begin{IEEEbiography}[{\includegraphics[width=1in,height=1.25in,clip,keepaspectratio]{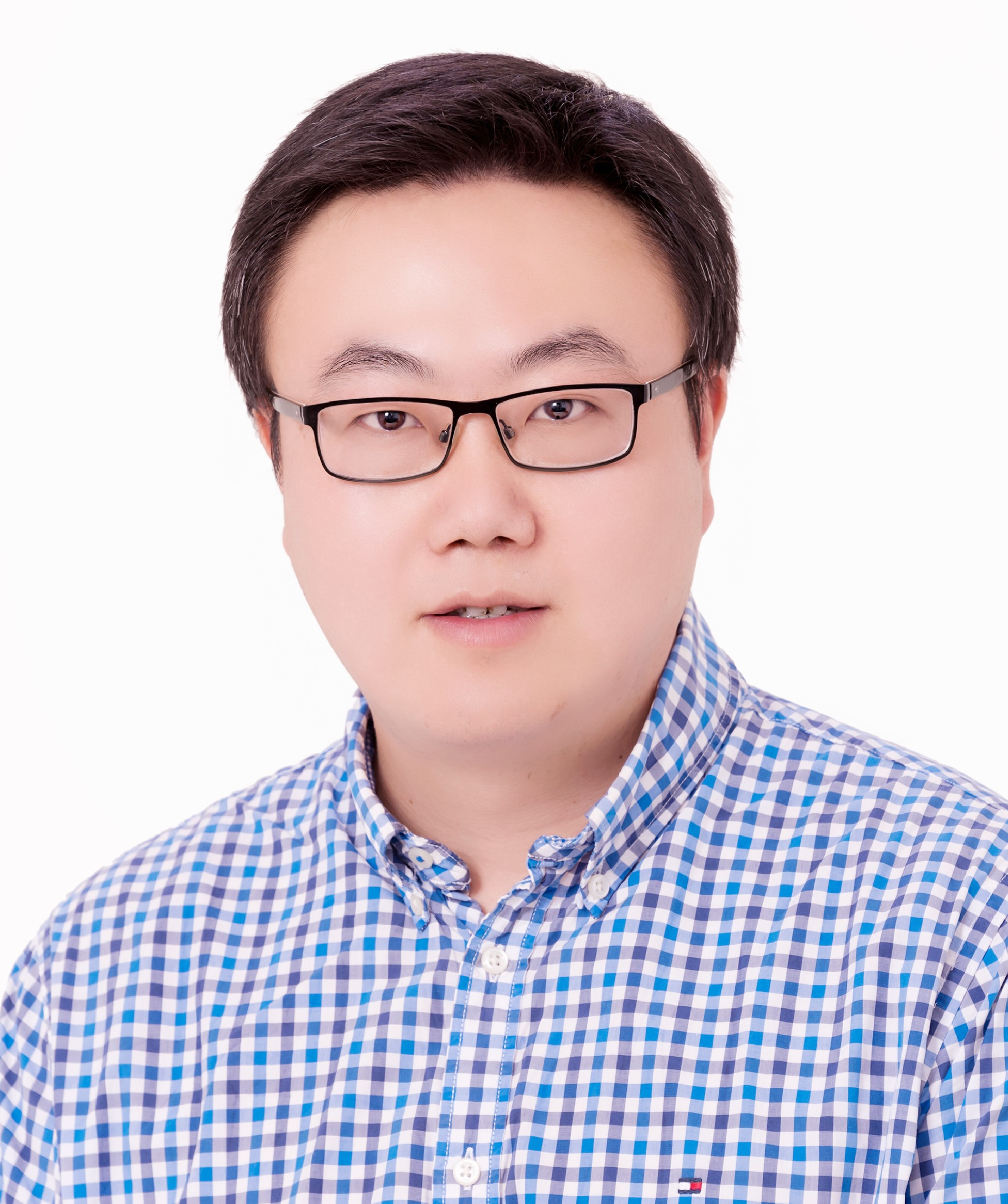}}]{Jiang Xue}(Senior Member, IEEE)
 		received the B.S. degree in Information and Computing Science from the Xi'an Jiaotong
 		University, Xi'an, China, in 2005, the M.S. degrees in Applied Mathematics from Lanzhou University, China and Uppsala University, Sweden, in 2008 and 2009, respectively. Prof. J. Xue received the Ph.D.degree in Electrical and Electronic Engineering from ECIT, the Queen's University of Belfast, U.K., in 2012. From 2013 to 2017, He was a Research Fellow with the University of Edinburgh, U.K.. Since 2017, Prof. J. Xue is with the National Engineering Laboratory for Big Data Analytics, Xi'an International Academy for Mathematics and Mathematical Technology, School of Mathematics and Statistics, Xi'an Jiaotong University, Pengcheng Lab, China, and he is supported by the
 		`Zhongying Young Scholars project'. His main interests include the machine
 		learning and wireless communication, performance analysis of multi-antenna systems, CSI estimation and prediction.
 	\end{IEEEbiography}

 	\begin{IEEEbiography}[{\includegraphics[width=1in,height=1.25in,clip,keepaspectratio]{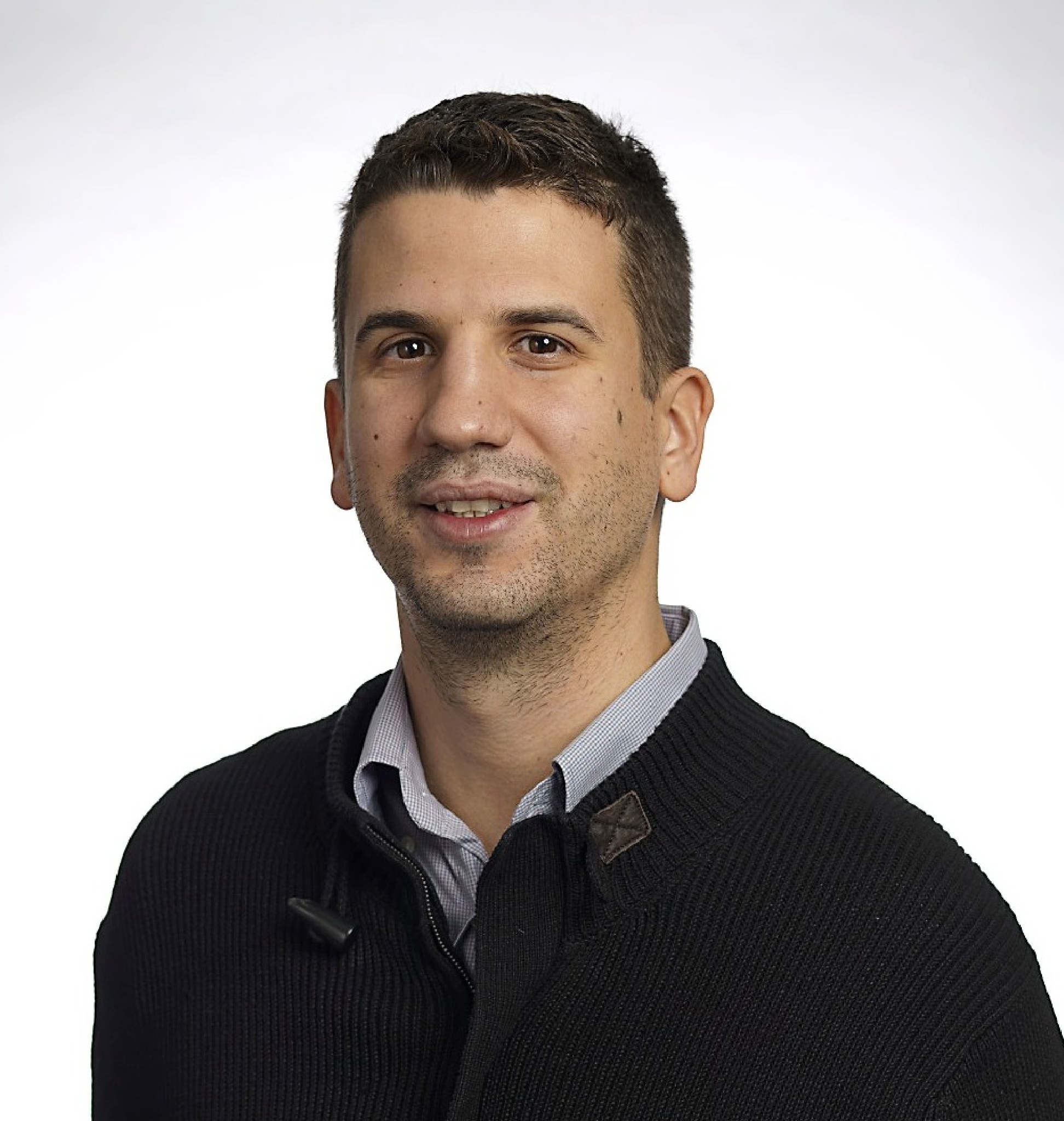}}]
 	{Michail Matthaiou}(Fellow, IEEE) was born in Thessaloniki, Greece in 1981. He obtained the Diploma degree (5 years) in Electrical and Computer Engineering from the Aristotle University of Thessaloniki, Greece in 2004. He then received the M.Sc. (with distinction) in Communication Systems and Signal Processing from the University of Bristol, U.K. and Ph.D. degrees from the University of Edinburgh, U.K. in 2005 and 2008, respectively. From September 2008 through May 2010, he was with the Institute for Circuit Theory and Signal Processing, Munich University of Technology (TUM), Germany working as a Postdoctoral Research Associate. He is currently a Professor of Communications Engineering and Signal Processing and Deputy Director of the Centre for Wireless Innovation (CWI) at Queen’s University Belfast, U.K. after holding an Assistant Professor position at Chalmers University of Technology, Sweden. His research interests span signal processing for wireless communications, beyond massive MIMO, intelligent reflecting surfaces, mm-wave/THz systems and deep learning for communications.

 	Dr. Matthaiou and his coauthors received the IEEE Communications Society (ComSoc) Leonard G. Abraham Prize in 2017. He currently holds the ERC Consolidator Grant BEATRICE (2021-2026) focused on the interface between information and electromagnetic theories. To date, he has received the prestigious 2023 Argo Network Innovation Award, the 2019 EURASIP Early Career Award and the 2018/2019 Royal Academy of Engineering/The Leverhulme Trust Senior Research Fellowship. His team was also the Grand Winner of the 2019 Mobile World Congress Challenge. He was the recipient of the 2011 IEEE ComSoc Best Young Researcher Award for the Europe, Middle East and Africa Region and a co-recipient of the 2006 IEEE Communications Chapter Project Prize for the best M.Sc. dissertation in the area of communications. He has co-authored papers that received best paper awards at the 2018 IEEE WCSP and 2014 IEEE ICC. In 2014, he received the Research Fund for International Young Scientists from the National Natural Science Foundation of China. He is currently the Editor-in-Chief of Elsevier Physical Communication, a Senior Editor for \textsc{IEEE Wireless Communications Letters} and \textsc{IEEE Signal Processing Magazine}, and an Associate Editor for \textsc{IEEE Transactions on Communications}. He is an IEEE Fellow.
 	%and an Associate Editor for the \textsc{IEEE JSAC Series on Machine Learning for Communications and Networks}. 
 	%In the past, he was an Associate Editor for the \textsc{IEEE Transactions on Communications} and Associate Editor/Senior Editor for \textsc{IEEE Communications Letters}. 
	 \end{IEEEbiography}
	
\end{document}